\pdfoutput=1
\RequirePackage{ifpdf}
\ifpdf
\documentclass[pdftex]{sigma}
\else
\documentclass{sigma}
\fi

\numberwithin{equation}{section}

\newcommand{\al}{\alpha}
\newcommand{\pa}{\partial}
\newcommand{\ep}{\epsilon}

\newcommand{\si}{\sigma}

\newcommand{\ta}{\tau}

\newcommand{\Om}{\Omega}
\newcommand{\om}{\omega}
\newcommand{\de}{\delta}

\newcommand{\De}{\Delta}

\newcommand{\rar}{\rightarrow}

\begin{document}
\allowdisplaybreaks

\renewcommand{\thefootnote}{$\star$}

\renewcommand{\PaperNumber}{003}

\FirstPageHeading

\ShortArticleName{From Quantum $A_N$ (Sutherland) to $E_8$ Trigonometric Model: Space-of-Orbits View}

\ArticleName{From Quantum $\boldsymbol{A_N}$ (Sutherland)\\ to $\boldsymbol{E_8}$ Trigonometric Model: Space-of-Orbits
View\footnote{This
paper is a~contribution to the Special Issue ``Superintegrability, Exact Solvability, and Special Functions''.
The full collection is available at
\href{http://www.emis.de/journals/SIGMA/SESSF2012.html}{http://www.emis.de/journals/SIGMA/SESSF2012.html}}}

\Author{Alexander~V.~TURBINER}

\AuthorNameForHeading{A.V.~Turbiner}

\Address{Instituto de Ciencias Nucleares, Universidad Nacional
Aut\'onoma de M\'exico, \\
Apartado Postal 70-543, 04510 M\'exico, D.F., Mexico}
\Email{\href{mailto:turbiner@nucleares.unam.mx}{turbiner@nucleares.unam.mx}}

\ArticleDates{Received September 21, 2012, in f\/inal form January 11, 2013; Published online January 17, 2013}

\Abstract{A number of af\/f\/ine-Weyl-invariant integrable and exactly-solvable quantum models with trigonometric
potentials is considered in the space of invariants (the space of orbits). These models are completely-integrable and
admit extra particular integrals.
All of them are characterized by (i) a~number of polynomial eigenfunctions and quadratic in quantum numbers eigenvalues
for exactly-solvable cases, (ii) a~factorization property for eigenfunctions, (iii) a~rational form of the potential
and the polynomial entries of the metric in the Laplace--Beltrami operator in terms of af\/f\/ine-Weyl (exponential)
invariants (the same holds for rational models when polynomial invariants are used instead of exponential ones), they
admit (iv) an algebraic form of the gauge-rotated Hamiltonian in the exponential invariants (in the space of orbits)
and (v) a~hidden algebraic structure. A hidden algebraic structure for $(A{-}B{-}C{-}D)$-models, both rational and
trigonometric, is related to the universal enveloping algebra $U_{gl_n}$. For the exceptional $(G{-}F{-}E)$-models, new,
inf\/inite-dimensional, f\/initely-generated algebras of dif\/ferential operators occur.
Special attention is given to the one-dimensional model with $BC_1\equiv(\mathbb{Z}_2)\oplus T$ symmetry. In
particular, the $BC_1$ origin of the so-called TTW model is revealed. This has led to a~new quasi-exactly solvable
model on the plane with the hidden algebra $sl(2)\oplus sl(2)$.}


\Classification{35P99; 47A15; 47A67; 47A75}

\renewcommand{\thefootnote}{\arabic{footnote}}
\setcounter{footnote}{0}

\section{Introduction}

In this article we attempt to overview our constructive knowledge of (quasi)-exactly-solvable potentials having the
form of a~meromorphic function in trigonometric variables. Any model with such a~potential is characterized by a
discrete symmetry group, and possesses an (in)f\/inite set of polynomial eigenfunctions in a~certain trigonometric
variables. In the case of exactly-solvable potentials an inf\/inite discrete spectra is quadratic in the quantum numbers.
All of these models are characterized by the appearance of a~hidden (Lie) algebraic structure. They do not admit a
separation of variables, they are completely-integrable and possess a~commutative algebra of integrals. So far, no
super-integrable models with trigonometric potentials are known, although all of them admit at least one particular
integral~\cite{Turbiner:2012}.

A similar overview of the rational models (with a~potential in the form of a~meromorphic function in Cartesian
coordinates) was given in~\cite{Turbiner:2011}. Unlike the trigonometric models the rational models do admit a
separation out of radial coordinate and hence, the emergence of integral of the second order leading to
super-integrability. For exactly-solvable rational models their eigenvalues depend on quantum numbers linearly, thus,
their spectrum is a~linear superposition of equidistant spectra.

Any spinless quantum system is characterized by the Hamiltonian
\begin{gather}
\label{H}
{\cal H}=-\De+V(x),\qquad x\in{\mathbb R}^d.
\end{gather}
The main problem of quantum mechanics is to f\/ind the spectrum the Hamiltonian looking for the Schr\"odinger equation
\begin{gather*}
{\cal H}\Psi(x)=E\Psi(x),\qquad\Psi(x)\in L^2\big({\mathbb R}^d\big).
\end{gather*}
in the Hilbert space. Since the Hamiltonian is a~dif\/ferential operator it can be represented as inf\/inite-dimensional
matrix. Thus, the solving the Schr\"odinger equation is equivalent to diagonalizing the inf\/inite-dimensional matrix.
It is a~transcendental problem: the characteristic polynomial is of inf\/inite order and it has inf\/initely-many roots. In
general, we do not know how to make such a~diagonalization explicitly. One can try to describe or construct quantum
system for which a~transcendental nature of~\eqref{H} degenerates (completely or partially) to algebraic one:
the roots of the characteristic polynomial (energies), some or all, can be found explicitly (algebraically). Usually,
in such a~situation one can indicate an analytic form of (some or all) eigenfunctions.
Such systems do exist and we call them {\it solvable}. If all energies are known they are called {\it
exactly-solvable} (ES), if only some number of them is known we call them {\it
quasi-exactly-solvable} (QES)~\cite{Turbiner:1988}.
Surprisingly, almost all such models the present author is familiar with, are provided by integrable systems emerging
from the Hamiltonian reduction method~\cite{OP} with real, continuous coupling constants. Sometimes, these models are
called the Calogero--Moser--Sutherland models. Every Hamiltonian has a~discrete symmetry~-- it is symmetric with
respect to af\/f\/ine Weyl group. Usually, the multi-dimensional Hamiltonians of the trigonometric models are of the form
\begin{gather}
\label{HES}
{\cal H}\ =\ \frac{1}{2}\sum_{k=1}^{N}
\left[-\frac{\pa^{2}}{\pa y_{k}^{2}}\right]\ +
\frac{\beta^2}{8}\sum_{\al \in R_{+}}
\nu_{|\al|}(\nu_{|\al|}-1)\frac{| \al|^{ 2}}
{\sin^2 \frac{\beta}{2} (\al\cdot y)},
\end{gather}
in the exactly-solvable case, where $R_+$ is a~set of positive roots in the root space $\De$ of dimen\-sion~$N$, $\beta$
is a~parameter and $\mu_{|\al|}$ are coupling constants which depend on the root length. For roots of the same length
the constants $\nu_{|\al|}$ are equal. Thus, the potential in~\eqref{HES} is a~superposition of the Weyl-invariant
functions, each def\/ined as a~sum over roots of the same length. The conf\/iguration space is the Weyl alcove. The ground
state wave function has a~form
\begin{gather}
\label{GSES}
\Psi_0 (y) = \prod_{\al\in R_+}
\left|\sin \frac{\beta}{2}(\al \cdot y)\right|^{\nu_{|\al|}}.
\end{gather}
The ground state energy has a~form $E_0=\beta^2{\ep}_0(\nu)$ and is known explicitly.

Let us take the Hamiltonian in ${\mathbb R}^d$ which is symmetric with respect to the (maximal) discrete group $G$. One can
construct invariants of $G$ using a~procedure of averaging some function over orbit(s). The $d$ linearly independent
invariants span a~linear space called the {\it space of orbits}.
These invariants are generating elements of the algebra of invariants. The main idea of this paper is to study the
Hamiltonian in a~space of orbits (space of invariants). Technically, it implies a~change of variables from the original
coordinates to the invariants. Conceptually, it means factoring out the discrete symmetry of the problem. It reveals
a~``primary'' operator of the system which being dressed by the discrete symmetry becomes the Hamiltonian.

We consider some models from the list of ones known so far.

\section{Solvable models}

\subsection{Generalities}

Many years ago, as the state-of-the-art, Sutherland found a~many-body exactly-solvable and integrable Hamiltonian
with trigonometric potential~\cite{Sutherland:1971}. A few years later the
Hamiltonian reduction method was introduced (for review and references see e.g.\ Olshanetsky--Perelo\-mov~\cite{OP}).
In this method an extended family of integrable and exactly-solvable Hamiltonians with trigonometric potentials,
associated with af\/f\/ine Weyl (Coxeter) symmetry, was found. The Sutherland model appeared as one of its representatives,
the
$A_N$ trigonometric model.

The idea of the Hamiltonian reduction method is beautiful:
\begin{itemize}\itemsep=0pt
\item Take a~simple group $G$.

\item Def\/ine the Laplace--Beltrami (invariant) operator on its symmetric space
(free motion).

\item Radial part of the Laplace--Beltrami operator is the Olshanetsky--Perelomov Hamiltonian relevant from physical
point of view. The emerging Hamiltonian is af\/f\/ine Weyl-symmetric,
it can be associated with root system, it is integrable with integrals given by the invariant operators of order higher
than two with a~property of solvability.
\end{itemize}

{\bf Trigonometric case.}
This case appears when the coordinates of the symmetric space are introduced in such a~way that a~negative-curvature
surface occurs. Emerging the Calogero--Moser--Sutherland--Olshanetsky--Perelomov Hamiltonian in the Cartesian
coordinates has the form~\eqref{HES} with the ground state given by~\eqref{GSES}. In the Hamiltonian reduction, the
parame\-ters~$\nu_{|\al|}$ of the Hamiltonian take
a set of discrete values, however, they can also be generalized to any real value without
loosing the property of integrability as well as solvability with the only
constraint being the existence of $L^2$-solutions of the corresponding
Schr\"odinger equation. The conf\/iguration space for~\eqref{HES} is the Weyl alcove.

The Hamiltonian~\eqref{HES} is completely-integrable: there exists a~commutative algebra of integrals (including the
Hamiltonian) of dimension which is equal to the dimension of the con\-f\/i\-gu\-ra\-tion space (for integrals, see
Oshima~\cite{Oshima} with explicit forms of those). The Hamiltonian~\eqref{HES} is invariant with respect to the af\/f\/ine
Weyl (Coxeter) group transformation, which is the discrete symmetry group of the corresponding root space, see
e.g.~\cite{OP}.

The Hamiltonian~\eqref{HES} has a~hidden (Lie)-algebraic structure. In order to reveal it (see~\cite{Boreskov:2001, BLT:2008, BTLG:2011,Brink:1997, VGT:2009,Rosenbaum:1998, RT:1995}) we need to
\begin{itemize}\itemsep=0pt
\item Gauge away the ground state eigenfunction making a~{\it similarity transformation}
$(\Psi_{0})^{-1}({\cal H}-E_0)\Psi_{0}=h$, then

\item If the state-of-art variables are introduced for trigonometric models $A_N$, $BC_N$, $G_2$ and $F_4$
(see~\cite{Boreskov:2001, Brink:1997,Rosenbaum:1998, RT:1995}, respectively), the Hamiltonian $h$ becomes
algebraic,
however,

\item It can be checked, which, in fact, looks evident, that parameterizing the space of orbits of the Weyl (Coxeter)
group by taking the {\it Weyl $($Coxeter$)$ fundamental trigonometric invariants},
\begin{gather}
\label{INV}
\ta_{a}^{(\Om)}(y;\beta)=\sum_{w\in\Om_a}e^{i\beta(w,y)},
\end{gather}
where $\Om_a$ is an orbit generated by {\it fundamental weight}
$w_a$, $a=1,2,\ldots,N$ ($N$ is the rank of the root system); $\vec{y}$ is $N$-dimensional auxiliary vector which
def\/ines the Cartesian coordinates, as coordinates we arrive at the conclusion that the state-of-the-art
variables~\cite{Boreskov:2001, Brink:1997,Rosenbaum:1998, RT:1995} coincide with~\eqref{INV}. It should be emphasized
that this fact was not clear to the authors of articles~\cite{Boreskov:2001, Brink:1997,Rosenbaum:1998, RT:1995}
including the present author.

From physical point of view, the expression~\eqref{INV} is a~Weyl-invariant non-linear superposition of plane waves
with momenta proportional to $\beta$.
\end{itemize}

The fundamental trigonometric invariants $\ta(\beta)$ taken as coordinates {\it always} lead to the gauge-rotated
trigonometric Hamiltonian $h$ in a~form of {\it algebraic} dif\/ferential operator with polynomial coef\/f\/icients. It is
proved by demonstration. It is worth emphasizing a~surprising fact that the period(s) of the invariants $\ta(\beta)$ is
half of the period(s) of the Hamiltonian~\eqref{HES} and the ground state function \eqref{GSES}. It seems correct (this
can be proved by demonstration) that the original Hamiltonian ${\cal H}$~\eqref{HES} written in terms of the
fundamental trigonometric invariants $\ta(\beta)$ takes the
form
\begin{gather}
\label{H-tau-rat}
{\cal H}(\ta)=-\De_g+V(\ta),
\end{gather}
where
\begin{gather*}
\De_g=\frac{1}{\sqrt{g}}\pa_{\ta_{i}}\sqrt{g}g^{ij}(\ta)\pa_{\ta_{j}},
\end{gather*}
is the Laplace--Beltrami operator with a~metric $g^{ij}(\ta)$ with polynomial in $\ta$ matrix elements, hence with
polynomial in $\tau$ coef\/f\/icient functions in front of the second derivatives, and with the property that the
coef\/f\/icient functions in front of the f\/irst derivatives are also polynomials in $\tau$; $V(\ta)$ is a~rational
function, see below e.g.~\eqref{HBC1-tau},~\eqref{HBC1-tau-qes}. The form \eqref{H-tau-rat} can be called the rational
form of the trigonometric model. It is evident that the similar rational form~\eqref{H-tau-rat} appears
for rational models when polynomial Weyl invariants are used as new coordinates to parameterize the space of orbits. In
turn, the gauge-rotated Hamiltonian $h$ in $\ta$-variables takes the form
\begin{gather*}
h(\ta)=-\De_g+\sum_{\al\in R_{+}}
\nu_{|\al|}\sum_{a=1,\ldots, N}C_a^{|\al|}(\ta)\pa_{\ta_{a}},
\end{gather*}
where $C_a(\ta)$ are polynomials in $\ta$, see below e.g.~\eqref{hBC1},~\eqref{hAN}. The same representation is valid
for the rational models.

\subsection[$A_1/BC_1$ case or trigonometric P\"oschl-Teller potential]{$\boldsymbol{A_1/BC_1}$ case or trigonometric
P\"oschl--Teller potential}

The $BC_1$ trigonometric Hamiltonian reads\footnote{Common factor $\frac{1}{2}$ is omitted.}
\begin{gather}
\label{HBC1}
{\cal H}_{BC_1}(x)=-\frac{d^2}{d{x}^2}
+
\frac{g_{2}\beta^2}{\sin^2\beta x}+
\frac{g_{3}\beta^2}{4\sin^2{\frac{\beta x}{2}}},
\end{gather}
where $\beta$, $g_2$, $g_3$ are parameters. Symmetry:  $(\mathbb{Z}_2)\oplus T$ (ref\/lections $x\rar-x$,
translation $x\rar x+2\pi/\beta$). As for the conf\/iguration space, it can be taken the
interval $[0,\frac{\pi}{\beta}]$. If $g_2=0$ the interval can be extended to
$[0,\frac{2\pi}{\beta}]$. At $g_3=0$ (or $g_2=0$) the Hamiltonian~\eqref{HBC1} degenerates to the $A_1$
trigonometric Hamiltonian, which describes the relative motion of two particle system on a~line.

The ground state for~\eqref{HBC1} reads
\begin{gather}
\label{psi_bc1}
\Psi_{0}=\big|\sin(\beta x)\big|^{\nu_{2}}\left|\sin\left({\frac{\beta}{2}{x}}\right)\right|^{\nu_3},\qquad
E_0=-\left(\nu_2+\frac{\nu_3}{2}\right)^2\beta^2,
\end{gather}
cf.~\eqref{GSES}, where $\nu_2$, $\nu_3$ are found from the relations
\begin{gather*}
g_{2}=\nu_{2}(\nu_{2}-1)>-\frac{1}{4},\qquad g_{3}=\nu_{3}(\nu_{3}+2\nu_{2}-1)>-\frac{1}{4}.
\end{gather*}
Note that if the parameters in~\eqref{HBC1} are related $g_{2}=\frac{g_{3}}{2}(\frac{g_{3}}{2}-1)$, the ground state
energy~\eqref{psi_bc1} reaches its maximal value, $E_0=0$.

Any eigenfunction has a~form $\Psi_{0}\varphi$, where $\varphi$ is a~polynomial in the $BC_1$ fundamental
trigonometric invariant $\ta(\beta)=\cos(\beta x)$ (see~\eqref{INV}). Hence, the ground state function $\Psi_{0}$
plays a~role of a~multiplicative factor.

The $BC_1$ trigonometric Hamiltonian~\eqref{HBC1} is easily related to the trigonometric P\"oschl--Teller (PT)
Hamiltonian
\begin{gather}
\label{HP-T}
{\cal H}_{\rm PT}=-\frac{d^2}{d{x}^2}
+\frac{\big(\al^{2}-\frac{1}{4}\big)\beta^2}{4\sin^2\frac{\beta
x}{2}}+\frac{\big(\gamma^{2}-\frac{1}{4}\big)\beta^2}{4\cos^2\frac{\beta x}{2}},
\end{gather}
where
\begin{gather*}
\al^2-\frac{1}{4}=g_2+g_3,\qquad\gamma^2-\frac{1}{4}=g_2.
\end{gather*}
Replacing in~\eqref{HP-T} $\beta\rar i\beta$, we arrive at the general hyperbolic P\"oschl--Teller Hamiltonian
\begin{gather*}
{\cal H}_{\rm PT}^{(h)}=-\frac{d^2}{d{x}^2}
+\frac{\big(\al^{2}-\frac{1}{4}\big)\beta^2}{4\sinh^2\frac{\beta
x}{2}}-\frac{\big(\gamma^{2}-\frac{1}{4}\big)\beta^2}{4\cosh^2\frac{\beta x}{2}},
\end{gather*}
while the one-soliton Hamiltonian appears at $\al^2=\frac{1}{4}$. In the case of the $BC_1$ trigonometric Hamiltonian
under the replacement $\beta\rar i\beta$ the $BC_1$ Hyperbolic Hamiltonian occurs.

Let us introduce a~new variable
\begin{gather}
\label{tau}
\ta=\cos(\beta x)
\end{gather}
(which is the $\frac{2\pi}{\beta}$-periodic, $BC_1$-Weyl invariant) in the $BC_1$ Hamiltonian~\eqref{HBC1}. It appears
that
\begin{gather}\label{HBC1-tau}
{\cal H}_{BC_1}(\ta)=-\De_g
+\frac{g_{2}}{2(1+\ta)}+\frac{(g_{2}+g_{3})}{2(1-\ta)},
\end{gather}
with amazingly simple meromorphic potential, where
\begin{gather*}
\De_g=\big(\ta^2-1\big)\frac{d^2}{d{\ta}^2}+\ta\frac{d}{d{\ta}}
\end{gather*}
is the f\/lat Laplace--Beltrami operator with metric $g^{11}=\big(\ta^2-1\big)$. Overall multiplicative factor $\beta^2$
in~\eqref{HBC1-tau} is dropped of\/f.
It can be called a~rational form of the $BC_1$ trigonometric Hamiltonian. The eigenvalue problem for~\eqref{HBC1-tau}
is considered on the interval $[-1,1]$.
It can be easily seen that the rational form for the $BC_1$ hyperbolic Hamiltonian is exactly the same as for
the $BC_1$ trigonometric Hamiltonian~(!) and is given by~\eqref{HBC1-tau}. However, the domain for the $BC_1$
hyperbolic Hamiltonian~\eqref{HBC1-tau} is $[1,\infty)$. In the hyperbolic case taking $\ta=\cosh{\beta
x}$ (cf.~\eqref{tau}) we obtain the same Hamiltonian~\eqref{HBC1-tau}. The ground state eigenfunction \eqref{psi_bc1}
in $\ta$ coordinate~\eqref{tau} becomes
\begin{gather}
\label{psi0-BC1T}
\Psi_0(\ta)=(1+\ta)^{\frac{\nu_2}{2}}(1-\ta)^{\frac{\nu_2+\nu_3}{2}}.
\end{gather}
At $\nu_2=1$ and $\nu_3=0$ it coincides to the Jacobian.

Now let us make a~gauge rotation
\begin{gather*}
h_{BC_1}=\frac{1}{\beta^2}\Psi_{0}^{-1}({\cal H}_{BC_1}-E_0)\Psi_{0},
\end{gather*}
with $\Psi_{0}$ given by~\eqref{psi_bc1} and write the result in the variable $\ta$.
After a~simple calculations it reads
\begin{gather}\label{hBC1}
h_{BC_1}(\ta)=\big(\ta^2-1\big)\frac{d^2}{d{\ta}^2}+[(2\nu_2+\nu_3+1)\ta+\nu_3]\frac{d}{d{\ta}},
\end{gather}
which is the algebraic form of the ${BC_1}$ Hamiltonian~\eqref{HBC1}. Its
eigenvalues are
\begin{gather}\label{EBC1}
\ep_p=p^2+(2\nu_2+\nu_3)p,
\qquad
p=0,1,2,\ldots,
\end{gather}
being quadratic in quantum number $p$, while the eigenfunctions are the Jacobi polynomials, $\varphi_p=
P_p^{(\nu_2+\nu_3-\frac{1}{2},\nu_2-\frac{1}{2})}(\ta)$.
Eventually, the explicit form of an eigenfunction of the Hamiltonian~\eqref{HBC1} is
\begin{gather*}
\Psi_p^{(BC_1)}
=P_p^{(\nu_2+\nu_3-\frac{1}{2},\nu_2-\frac{1}{2})}(\cos(\beta x))\big|\sin(\beta x)\big|^{\nu_{2}}
\left|\sin\left({\frac{\beta}{2}{x}}\right)\right|^{\nu_3},
\qquad
p=0,1,2,\ldots.
\end{gather*}

It can be easily checked that the gauge-rotated Hamiltonian $h_{BC_1}(\ta)$ has inf\/initely many f\/inite-dimensional
invariant subspaces
\begin{gather}
\label{P}
{\cal P}_{n}=\langle\ta^{p}\vert0\le p_1\le n\rangle,\qquad n=0,1,2,\ldots,
\end{gather}
hence, the inf\/inite f\/lag ${\cal P}$,
\begin{gather*}
{\cal P}_0\subset{\cal P}_1\subset{\cal P}_2\subset\cdots
\subset{\cal P}_n\subset\cdots\subset {\cal P},
\end{gather*}
with the characteristic vector $\vec f=(1)$ (see below), is preserved by $h_{BC_1}$. Thus, the eigenfunctions
of $h_{BC_1}$ are elements of the f\/lag ${\cal P}$.
Any subspace ${\cal P}_n$ contains $(n+1)$ eigenfunctions which is equal to
$\dim{\cal P}_n$.

Take the algebra $gl_2$ in $(n+1)$-dimensional representation realized by the
f\/irst order dif\/feren\-tial operators
\begin{gather}
J^-=\frac{d}{d\ta},
\qquad
J^0_n=\ta\frac{d}{d\ta}-n,\qquad T^0=1,\qquad
J^+_n={\ta}^2\frac{d}{d\ta}-n t=\ta J^0_n,\label{gl2}
\end{gather}
where $n=0,1,\ldots$ and $T^0$ is the central element. Its f\/inite-dimensional representation space is the space of
polynomials ${\cal P}_{n}$~\eqref{P}. Hence, the f\/inite-dimensional invariant subspaces of the Hamiltonian $h_{BC_1}$
coincide with the f\/inite-dimensional representation spaces of $gl_2$~\eqref{gl2} for $n=0,1,2,\ldots$. It immediately
implies that the algebra $gl_2$ is the hidden algebra of the ${BC_1}$ trigonometric Hamiltonian~-- it can be written in
terms of $gl_2$ generators~\eqref{gl2}
\begin{gather}
\label{BC1-hidden}
h_{BC_1}=J^0J^0-J^-J^-+(2\nu_2+\nu_3+1)J^0+\nu_3J^-,
\end{gather}
where $J^0\equiv J^0_0$, $J^-\equiv J^-_0$. Thus, the Hamiltonian $h_{BC_1}$ is an element of the universal
enveloping algebra $U_{gl_2}$.

Among the generators of the algebra $gl_2$~\eqref{gl2} there is the Euler-Cartan operator,
\begin{gather*}
J^0_n=\ta\frac{d}{d\ta}-n,
\end{gather*}
which has zero grading; it maps a~monomial in $\ta$ to itself. It def\/ines the highest weight vector. This generator
allows us to construct a~particular integral~-- $\pi$-integral of zero grading of the $(n+1)$th
order (see~\cite{Turbiner:2012}) $i_{\rm par}^{(n)}(\ta)$: its commutator with $h_{BC_1}$ vanishes on a~subspace. If
\begin{gather}\label{pi1}
i_{\rm par}^{(n)}(\ta)=\prod_{j=0}^n\big(J^0_n+j\big),
\end{gather}
then
\begin{gather*}
\big[h_{BC_1}(\ta),i_{\rm par}^{(n)}(\ta)\big]: \ {\cal P}_{n}\mapsto0.
\end{gather*}
Making the gauge rotation of the $\pi$-integral~\eqref{pi1} with $\Psi_{0}^{-1}(\ta)$ given by \eqref{psi_bc1} and
changing variables $\ta$ back to the Cartesian coordinate we arrive at the quantum $\pi$-integral acting in the Hilbert
space,
\begin{gather*}
{\cal I}_{{\rm par},BC_1}^{(n)}(x)=\left.\Psi_{0}(\ta)i_{\rm par}^{(n)}(\ta)\Psi_{0}^{-1}(\ta)\right|_{\ta\rar x}.
\end{gather*}
Under such a~gauge transformation the triangular space of polynomials ${\cal P}_{n}$ becomes the space
\begin{gather*}
{\cal V}_{n}=\Psi_0{\cal P}_{n}.
\end{gather*}
The Hamiltonian ${\cal H}_{BC_1}(x)$ commutes with ${\cal I}_{{\rm par},BC_1}^{(n)}(x)$ over
this space
\begin{gather*}
\big[{\cal H}_{BC_1}(x),{\cal I}_{{\rm par},BC_1}^{(n)}(x)\big]: \
{\cal V}^{(N-1)}_{n}\mapsto0.
\end{gather*}
Any eigenfunction $\Psi\in{\cal V}_{n}$ is zero mode of the $\pi$-integral ${\cal I}_{{\rm par},BC_1}^{(n)}(x)$.

It is worth noting a~connection of the $BC_1$ trigonometric model with the so-called Tremblay--Turbiner--Winternitz (TTW)
model~\cite{TTW:2009} and, in particular, with the $I_{2}(k)$ rational model (see e.g.~\cite{Turbiner:2011}). In order
to see it let us
combine the~$BC_1$ trigonometric Hamiltonian ${\cal H}_{BC_1}(\phi)$~\eqref{HBC1} as the angular part and the radial
part of two-dimensional spherical-symmetrical
harmonic oscillator Hamiltonian as the radial part forming the 2D Hamiltonian
\begin{gather}
\label{H_TTW}
{\cal H}_{\rm TTW}(r,\phi;\om,\nu_2,\nu_3,\beta)=-\pa_r^2-\frac{1}{r}\pa_r+\om^2r^2+\frac{{\cal
H}_{BC_1}(\phi)}{r^2},
\end{gather}
which is nothing but the Hamiltonian of the TTW model~\cite{TTW:2009}. If $\beta=k$ is integer, this Hamiltonian
corresponds to the $I_{2}(k)$ rational model~\cite{OP}.
Since the both Hamiltonians are exactly-solvable, the TTW model is also exactly-solvable but with spectra of
two-dimensional anisotropic~(!) harmonic oscillator with frequency ratio $1:\beta$. Any eigenfunction
of~\eqref{H_TTW} has the form of a~polynomial $p(r^2,\cos(\beta\phi))$ in variables $r^2$ and $\cos(\beta\phi)$
multiplied by a~ground state function
\begin{gather}
\label{psi0_TTW}
\Psi_0^{\rm (TTW)}(r,\phi)=
r^{(\nu_2+\nu_3)\beta}\big|\sin(\beta\phi)\big|^{\nu_{2}}
\left|\sin\left({\frac{\beta}{2}{\phi}}\right)\right|^{\nu_3}e^{-\frac{\om r^2}{2}},
\end{gather}
namely,
\begin{gather*}
\Psi^{\rm (TTW)}(r,\phi)=p\left(r^2,\cos(\beta\phi)\right)\Psi_0^{\rm (TTW)}(r,\phi).
\end{gather*}

If in the construction~\eqref{H_TTW} instead of two-dimensional radial harmonic oscillator, the radial Hamiltonian of
the sextic QES $2D$ central potential (see e.g.~\cite{Turbiner:1988}) is taken, the quasi-exactly-solvable extension of
the TTW model occurs~\cite{TTW:2009}
\begin{gather}
{\cal H}_{\rm TTW}^{\rm (qes)}(r,n;\phi;\om,\nu_2,\nu_3,\beta,a)=-\pa_r^2-\frac{1}{r}\pa_r+a^2r^6+2a\om r^4
\nonumber\\
\hphantom{{\cal H}_{\rm TTW}^{\rm (qes)}(r,n;\phi;\om,\nu_2,\nu_3,\beta,a)=}{}
+\left[\om^2-2a(2n+2+\beta(\nu_2+\nu_3))\right]r^2+\frac{{\cal H}_{BC_1}(\phi)}{r^2},\label{H_qes}
\end{gather}
cf.~\eqref{H_TTW}, here $n$ is non-negative integer and $a>0$ is a~parameter.
In this Hamiltonian a~f\/inite number of eigenstates can be found explicitly (algebraically). Their eigenfunctions have
the form of a~polynomial
$p(r^2,\cos(\beta\phi))$ of degree $n$ in $r^2$ multiplied by a~factor
\begin{gather}
\label{psi0_qes}
\Psi_0^{\rm (qes,TTW)}=
r^{(\nu_2+\nu_3)\beta}\big|\sin(\beta\phi)\big|^{\nu_{2}}\left|\sin\left({\frac{\beta}{2}{\phi}}\right)\right|^{\nu_3}e^{-\frac{\om
r^2}{2}-\frac{a r^4}{4}},
\end{gather}
cf.~\eqref{psi0_TTW},
namely,
\begin{gather*}
\Psi_{\rm alg}^{\rm (qes,TTW)}=p\left(r^2,\cos(\beta\phi)\right)\Psi_0^{\rm (qes,TTW)}.
\end{gather*}
The factor~\eqref{psi0_qes} is the ground state eigenfunction of the Hamiltonian~\eqref{H_qes} at $n=0$.
If $\beta$ is equal to non-negative integer $k$, a~polynomial $p (r^2,\cos(\beta\phi) )$ belongs to the space ${\cal
P}_{(1,k)}$ with the characteristic vector $\vec f=(1,k)$, see below.

\subsection[Quasi-exactly-solvable $BC_1$ case (or QES trigonometric P\"oschl-Teller potential)]{Quasi-exactly-solvable $\boldsymbol{BC_1}$ case\\ (or QES trigonometric P\"oschl--Teller potential)}

The Hamiltonian $h_{BC_1}(\ta)$~\eqref{hBC1} is $gl(2)$-Lie-algebraic operator \eqref{BC1-hidden} which has
inf\/initely-many f\/inite-dimensional invariant subspaces in polynomials~\eqref{P}. By adding
to $h_{BC_1}(\ta)$~\eqref{hBC1} the operator
\begin{gather*}
\de h^{\rm (qes)}(\ta)=2b\big(\ta^2-1\big)\frac{d}{d\ta}-2b n\ta+2b\left(n+\nu_2+\nu_3+\frac{1}{2}\right),
\end{gather*}
where $b$ is a~parameter and $n$ is non-negative integer, as a~result
we get the operator
\begin{gather}\label{hqes_BC1}
h_{BC_1}^{\rm (qes)}(\ta)=h_{BC_1}+\de h^{\rm (qes)},
\end{gather}
which has a~single f\/inite-dimensional invariant subspace
\begin{gather*}
{\cal P}_n=\langle\ta^p\,|\,0\leq p\leq n\rangle,
\end{gather*}
of the dimension $(n+1)$. Hence, this operator is quasi-exactly-solvable - it can be written in terms of $gl_2$
generators in $(n+1)$-dimensional representation~\eqref{gl2},
\begin{gather*}
h_{BC_1}^{\rm (qes)}=J^0_n J^0_n-J^-J^--2b J^+_n+(2n+2\nu_2+\nu_3+1)J^0_n+(\nu_3-2b)J^-
\\
\phantom{h_{BC_1}^{\rm (qes)}=}{}
+n(n+2\nu_2+\nu_3+1).
\end{gather*}

Making the gauge rotation of~\eqref{hqes_BC1} with
\begin{gather*}
\tilde\Psi_0=e^{-\frac{\nu_2+\nu_3}{2}\log(1-\ta)-\frac{\nu_2}{2}\log(1+\ta)}e^{b\ta}
\end{gather*}
and the change of variable $\ta=\cos(\beta x)$ we arrive at the $BC_1$-trigonometric QES
Hamiltonian~\cite{Turbiner:1988}
\begin{gather}
{\cal H}_{BC_1}^{\rm (qes)}(x)=-\frac{d^2}{d{x}^2}+
\frac{\nu_{2}(\nu_{2}-1)\beta^2}{\sin^2\beta x}+
\frac{\nu_{3}(\nu_{3}+2\nu_{2}-1)\beta^2}{4\sin^2{\frac{\beta x}{2}}}+
b^2\beta^2{\sin^2\beta x}
\nonumber\\
\phantom{{\cal H}_{BC_1}^{\rm (qes)}(x)=}{}
+2b\beta^2(2n+2\nu_{2}+\nu_{3}+1)\sin^2{\frac{\beta x}{2}},\label{BC1_qest}
\end{gather}
cf.~\eqref{HBC1}, where $b$, $\nu_2$, $\nu_3$, $\beta$ are parameters, $n$ is non-negative integer.
In $\ta$-variable~\eqref{tau} the $BC_1$-trigonometric QES Hamiltonian appears in rational form
\begin{gather}
\label{HBC1-tau-qes}
{\cal H}_{BC_1}^{\rm (qes)}(\ta) = -\De_g+\frac{g_{2}}{\big(1\!-\!\ta^2\big)}+
\frac{g_{3}}{2(1\!-\!\ta)}+b^2\big(1\!-\!\ta^2\big)+b(2n+2\nu_{2}+\nu_{3}+1)(1\!-\!\ta),
\end{gather}
cf.~\eqref{HBC1-tau}, where $\De_g= (\ta^2-1 )\frac{d^2}{d{\ta}^2}+\ta\frac{d}{d{\ta}}$ is the f\/lat
Laplace--Beltrami operator with metric $g^{11}= (\ta^2-1 )$. Overall multiplicative factor $\beta^2$
in~\eqref{BC1_qest} is dropped of\/f.

In the Hamiltonian~\eqref{BC1_qest} the $(n+1)$ eigenfunctions are of a~form
\begin{gather*}
P_n(\cos(\beta x))\big|\sin(\beta x)\big|^{\nu_{2}}
\left|\sin\left({\frac{\beta}{2}{x}}\right)\right|^{\nu_3}e^{-b\cos(\beta x)},
\end{gather*}
where $P_n(\ta)$ is a~polynomial of degree $n$, they can be found by algebraic means. It is evident
that $i_{\rm par}^{(n)}(\ta)$~\eqref{pi1} remains the particular integral~-- $\pi$-integral of the $BC_1$-trigonometric QES
Hamiltonian~\eqref{hqes_BC1} (see~\cite{Turbiner:2012})
\begin{gather*}
\big[h_{BC_1}^{\rm (qes)}(\ta),i_{\rm par}^{(n)}(\ta)\big]: \ {\cal P}_{n}\mapsto0.
\end{gather*}

Interestingly, the $BC_1$-trigonometric QES Hamiltonian~\eqref{BC1_qest} degenerates to
the so-called Magnus--Winkler (MW) Hamiltonian or, in other words, to the QES Lame Hamiltonian (see
e.g.~\cite{Turbiner:1988})
\begin{gather*}
{\cal H}_{BC_1}^{\rm (qes)}=-\frac{d^2}{d{x}^2}
+b^2\beta^2{\sin^2\beta x}+
2b\beta^2(2n+\nu+1)\sin^2{\frac{\beta x}{2}},
\end{gather*}
where $\nu=0,1$.

For $\nu=0$ and given $n$ there exist two families of eigenfunctions
\begin{gather*}
\varphi^{(0,+)}_{n,i}=P_n(\cos(\beta x))e^{-b\cos(\beta x)},\qquad i=0,1,\ldots,n,
\\
\varphi^{(0,-)}_{n-1,i}=P_{n-1}(\cos(\beta x))\sin(\beta x)e^{-b\cos(\beta x)},\qquad i=0,1,\ldots,n-1,
\end{gather*}
which correspond to periodic (anti-periodic) boundary conditions, correspondingly. These eigenfunctions describe
lower (upper) edges of Brillouin zones, respectively. Polynomial factors in~$\varphi^{(0,+)}_{n,i}$
and~$\varphi^{(0,-)}_{n-1,i}$ are eigenfunctions of
\begin{gather*}
h_{BC_1}^{({\rm qes},0,+)}=J^0_n J^0_n-J^-J^--2b J^+_n+(2n+1)J^0_n
-2b J^-+n(n+1),
\\
h_{BC_1}^{({\rm qes},0,-)}=J^0_{n-1}J^0_{n-1}-J^-J^--2b J^+_{n-1}+(2n+1)J^0_{n-1}-2b J^-+
n(n+2),
\end{gather*}
respectively (see~\eqref{gl2}).

For $\nu=1$ and given $n$ there also exist two families of eigenfunctions
\begin{gather*}
\varphi^{(1,-)}_{n,i}=P_n(\cos(\beta x))\sin\left({\frac{\beta}{2}{x}}\right)
e^{-b\cos(\beta x)},\qquad i=0,1,\ldots,n,
\\
\varphi^{(1,+)}_{n,i}=P_{n}(\cos(\beta x))\cos\left({\frac{\beta}{2}{x}}\right)e^{-b\cos(\beta x)},\qquad i=0,1,\ldots,n,
\end{gather*}
which correspond to (anti)-periodic boundary conditions, correspondingly. These eigenfunctions describe upper (lower)
edges of Brillouin zones, respectively. Polynomial factors in~$\varphi^{(1,-)}_{n,i}$ and~$\varphi^{(1,+)}_{n,i}$ are
eigenfunctions of
\begin{gather*}
h_{BC_1}^{({\rm qes},1,-)}=J^0_n J^0_n-J^-J^--2b J^+_n+2(n+1)J^0_n+
(1-2b)J^-+n(n+2),
\\
h_{BC_1}^{({\rm qes},1,+)}=J^0_n J^0_n-J^-J^--2b J^+_n+2(n+1)J^0_n-
(1+2b)J^-+n(n+2),
\end{gather*}
respectively (see~\eqref{gl2}).

If in a~construction~\eqref{H_TTW} to obtain the TTW model we replace the $BC_1$-trigonometric Hamiltonian ${\cal
H}_{BC_1}(\phi)$~\eqref{HBC1} by the $BC_1$-trigonometric QES Hamiltonian ${\cal
H}_{BC_1}^{\rm (qes)}(\phi)$~\eqref{hqes_BC1}
\begin{gather*}
{\cal H}_{\rm TTW}^{\rm (qes)}(r,\phi;\om,\nu_2,\nu_3,\beta)=-\pa_r^2-\frac{1}{r}\pa_r+\om^2
r^2+\frac{{\cal H}_{BC_1}^{\rm (qes)}(\phi)}{r^2},
\end{gather*}
a new quasi-exactly-solvable extension of the TTW model is obtained
\begin{gather*}
\tilde{\cal H}_{\rm TTW}^{\rm (qes)}(r;\phi,m;\om,\nu_2,\nu_3,\beta,b)=
-\De^{(2)}+\om^2r^2+
\frac{\nu_{2}(\nu_{2}-1)\beta^2}{r^2\sin^2\beta\phi}+
\frac{\nu_{3}(\nu_{3}+2\nu_{2}-1)\beta^2}{4r^2\sin^2{\frac{\beta\phi}{2}}}
\\ 
\qquad
{}+\frac{b^2\beta^2{\sin^2\beta\phi}}{r^2}+
\frac{2b\beta^2(2m+2\nu_{2}+\nu_{3}+1)\sin^2{\frac{\beta\phi}{2}}}{r^2},
\end{gather*}
cf.~\eqref{H_TTW},
where $\De^{(2)}$ is $2D$ Laplacian, $b$, $\nu_2$, $\nu_3$, $\beta$ are parameters, $m$ is non-negative integer.

If in the construction~\eqref{H_TTW} instead of two-dimensional radial harmonic oscillator, the radial Hamiltonian of
the sextic QES $2D$ radial potential~\cite{Turbiner:1988} is taken and the $BC_1$-trigonometric Hamiltonian ${\cal
H}_{BC_1}(\phi)$~\eqref{HBC1} is replaced by the $BC_1$-trigonometric QES Hamiltonian ${\cal
H}_{BC_1}^{\rm (qes)}(\phi)$ \eqref{hqes_BC1} the most general quasi-exactly-solvable extension of the TTW model occurs
\begin{gather}
\hat{\cal H}_{\rm TTW}^{\rm (qes)}(r,n;\phi,m;\om,\nu_2,\nu_3,\beta,a,b)=
-\De^{(2)}+a^2r^6+2a\om r^4
\nonumber\\
\qquad
{}+\left[\om^2-2a(2n+2+\beta(\nu_2+\nu_3))\right]r^2+
\frac{\nu_{2}(\nu_{2}-1)\beta^2}{r^2\sin^2\beta\phi}+
\frac{\nu_{3}(\nu_{3}+2\nu_{2}-1)\beta^2}{4r^2\sin^2{\frac{\beta\phi}{2}}}
\nonumber\\
\qquad
{}+\frac{b^2\beta^2{\sin^2\beta\phi}}{r^2}+
\frac{2b\beta^2(2m+2\nu_{2}+\nu_{3}+1)\sin^2{\frac{\beta\phi}{2}}}{r^2},\label{H_qesttt}
\end{gather}
where $n$, $m$ is non-negative integer and $a>0$, $b$ are parameters.
In this Hamiltonian a~f\/inite number of eigenstates can be found explicitly (algebraically). Their eigenfunctions have
the form of a~polynomial
$p\left(r^2,\cos(\beta\phi)\right)$ of degree $n$ in $r^2$ and of degree $m$ in $\cos(\beta x)$ multiplied by a~factor
\begin{gather}
\label{psi0_qes-TTW}
\hat\Psi_0^{\rm (qes,TTW)}(r,\phi)=
r^{(\nu_2+\nu_3)\beta}\big|\sin(\beta\phi)\big|^{\nu_{2}}
\left|\sin\left({\frac{\beta}{2}{\phi}}\right)\right|^{\nu_3}e^{-\frac{\om
r^2}{2}-\frac{a r^4}{4}-b\cos(\beta\phi)},
\end{gather}
cf.~\eqref{psi0_TTW}, namely,
\begin{gather*}
\hat\Psi_{\rm alg}^{\rm (qes,TTW)}(r,\phi)=p\left(r^2,\cos(\beta\phi)\right)\hat\Psi_0^{\rm (qes,TTW)}(r,\phi).
\end{gather*}
The factor~\eqref{psi0_qes-TTW} becomes the ground state eigenfunction of the Hamiltonian
\eqref{H_qesttt} at \mbox{$n\!=\!m\!=\!0$}.

\subsection[Case $A_{N-1}$]{Case $\boldsymbol{A_{N-1}}$}

\begin{figure}[t]\centering
\includegraphics[angle=-90, width=0.30\textwidth]{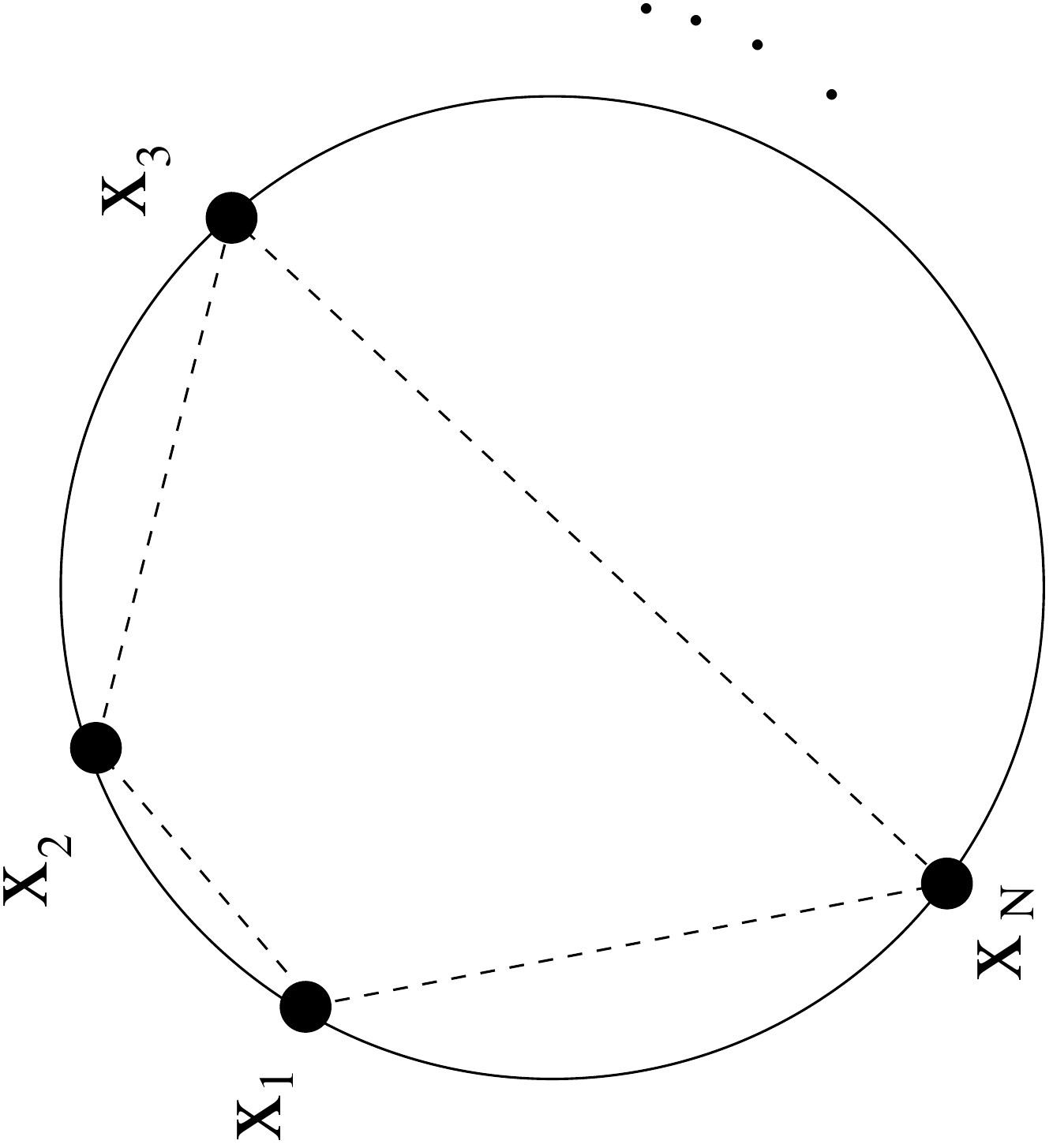}
\caption{$N$-body Sutherland model.}\label{Fig1}
\end{figure}

This is the celebrated Sutherland model ($A_{N-1}$ trigonometric model) which was found in~\cite{Sutherland:1971}. It
describes~$N$ identical particles on a~circle (see Fig.~\ref{Fig1}) with singular pairwise interaction $\propto\frac{1}{h^2}$
where~$h$ is the horde.
The Hamiltonian is
\begin{gather}
\label{HAN}
{\cal H}_{\rm Suth}=-\frac{1}{2}\sum_{k=1}^{N}\frac{\pa^{2}}{\pa x_{k}^{2}}
+\frac{g\beta^2}{4}\sum_{k<l}^{N}\frac{1}{\sin^{2}\big(\frac{\beta}{2}(x_{k}-x_{l})\big)},
\end{gather}
where $g$ is the coupling constant and $\beta$ is a~parameter. The symmetry of the system is $S_N\oplus
T\oplus\mathbb{Z}_2$ (permutations $x_i\rar x_j$, translation $x_i\rar x_i+2\pi/\beta$ and
all $x_i\rar-x_i$). The ground state of the Hamiltonian~\eqref{HAN} reads
\begin{gather}
\label{psi_suth}
\Psi_{0}(x)=\prod_{i<j}\left|\sin^2\left(\frac{\beta}{2}(x_{i}-x_{j})\right)\right|^{\nu},
\qquad
g=\nu(\nu-1)\geq-\frac{1}{4}
\end{gather}
(cf.~\eqref{GSES}).
Let us make the gauge rotation
\begin{gather*}
h_{\rm Suth}=\frac{2}{\beta^2}\Psi_{0}^{-1}({\cal H}_{\rm Suth}-E_0)\Psi_{0},
\end{gather*}
where $E_0$ is the ground state energy. Then introduce center-of-mass variables
\begin{gather*}
Y=\sum x_i,\qquad y_i=x_i-\frac{1}{N}Y,\qquad i=1,\ldots,N,
\end{gather*}
here $\sum\limits^N_{i=1}y_i=0$, and then new permutationally-symmetric, translationally-invariant, periodic relative
variables~\cite{RT:1995}
\begin{gather}
\label{cms}
(x_1,x_2,\ldots x_N)\rightarrow\big(Y, \ \ta_n(x)=\si_n\big(e^{i\beta y(x)}\big)\,\big|\,
{n=1,2,3,\ldots,(N-1)}\big),
\end{gather}
where
\begin{gather*}
\si_{k}(x)=\sum_{i_{1}<i_{2}<\cdots<i_{k}}
x_{i_{1}}x_{i_{2}}\cdots x_{i_{k}},
\qquad
\si_{k}(-x)=(-)^k\si_{k}(x),
\end{gather*}
are elementary symmetric polynomials, and
\begin{gather*}
\ta_0=\ta_N(x)=1,\qquad \ta_k(x)=0,\qquad k<0 \quad \text{or}\quad  k>N.
\end{gather*}
The ground state function~\eqref{psi_suth} in $\ta$-variables takes a~form of a~polynomial in some power, e.g.
\begin{gather*}
\Psi_{0}^{(A_2)}(x)=\left(4\ta_1^3+4\ta_2^3-18\ta_1\ta_2-\ta_1^2\ta_2^2+27\right)^{\frac{\nu}{2}}.
\end{gather*}

{\sloppy After the center-of-mass separation, the gauge rotated Hamiltonian takes the algebraic form~\cite{RT:1995}
\begin{gather}\label{hAN}
h_{\rm Suth}=\sum_{i,j=1}^{N-1}{\cal A}_{ij}(\ta)\frac{\pa^2}{\pa{\ta_i}\pa
{\ta_j}}+\sum_{i=1}^{N-1}{\cal B}_i(\ta)\frac{\pa}{\pa\ta_i},
\end{gather}
where
\begin{gather*}
{\cal A}_{ij}=\frac{(N-i)j}{N}\ta_{i}\ta_j+
\sum_{{l\geq}{\max(1,j-i)}}(j-i-2l)\ta_{i+l}\ta_{j-l},
\qquad
{\cal B}_i=\left(\frac{1}{N}+\nu\right)i(N-i)\ta_{i},
\end{gather*}
Eigenvalues of the gauge-rotated Hamiltonian~\eqref{hAN} are
\begin{gather*}
N\ep_{\{p\}}=\nu N\sum_{i=1}^{N-1}i(N-i)p_i+
\sum_{i,j=1}^{N-1}{(N-i)j}p_ip_j,
\end{gather*}
being quadratic in quantum numbers $\{p_1,p_2,\ldots, p_{(N-1)}\}$ where $p_1,p_2,\ldots, p_{(N-1)}=0,1,2,\ldots$.

}

It is easy to check that the gauge-rotated Hamiltonian $h_{\rm Suth}$ has inf\/initely many f\/inite-dimensional invariant
subspaces
\begin{gather}
\label{PN}
{\cal P}_n^{(N-1)}=\big\langle{\ta_{1}}^{p_1}
{\ta_{2}}^{p_2}\cdots
{\ta_{(N-1)}}^{p_{N-1}}
\,\vert\,0\le\sum p_i\le n\big\rangle.
\end{gather}
where $n=0,1,2,\ldots$. As a~function of $n$ the spaces ${\cal P}_n^{(N-1)}$ form the inf\/inite f\/lag (see below).

\subsubsection[The $gl_{d+1}$-algebra acting by 1st order differential operators in ${\mathbb R}^d$]{The $\boldsymbol{gl_{d+1}}$-algebra acting by 1st order dif\/ferential operators in $\boldsymbol{{\mathbb R}^d}$}

It can be checked by the direct calculation that the $gl_{d+1}$ algebra realized by the f\/irst-order dif\/ferential
operators acting in ${\mathbb R}^d$ in the representation given by the Young tableaux as a
row $(n,\underbrace{0,0,\ldots,0}_{d-1})$ has a~form
\begin{gather}
{\cal J}_i^-=\frac{\pa}{\pa\ta_i},\quad
i=1,2,\ldots, d,\qquad
{{\cal J}_{ij}}^0=
\ta_i\frac{\pa}{\pa\ta_j},\quad i,j=1,2,\ldots, d,
\nonumber\\
{\cal J}^0=\sum_{i=1}^{d}\ta_i\frac{\pa}{\pa\ta_i}-n,
\qquad
{\cal J}_i^+=\ta_i{\cal J}^0=
\ta_i\left(\sum_{j=1}^{d}\ta_j\frac{\pa}{\pa\ta_j}-n\right),
\quad i=1,2,\ldots, d, \label{gln}
\end{gather}
where $n$ is an arbitrary number. The total number of generators is $(d+1)^2$.
If $n$ takes the integer values, $n=0,1,2,\ldots$, the f\/inite-dimensional irreps occur
\begin{gather*}
{\cal P}_n^{(d)}=\big\langle{\ta_{1}}^{p_1}
{\ta_{2}}^{p_2}\cdots
{\ta_{d}}^{p_{d}}
\, \vert\, 0\le\sum p_i\le n\big\rangle
\end{gather*}
(cf.~\eqref{PN}).
It is a~common invariant subspace for~\eqref{gln}. The spaces ${\cal P}_n^{(d)}$ at $n=0,1,2,\ldots$ can be ordered
\begin{gather}
\label{flag}
{\cal P}_0^{(d)}\subset{\cal P}_1^{(d)}\subset{\cal P}_2^{(d)}\subset\cdots
\subset{\cal P}_n^{(d)}\subset\cdots\subset {\cal P}^{(d)}.
\end{gather}
Such a~nested construction is called {\em infinite flag $($filtration$)$}
${\cal P}^{(d)}$. It is worth noting that the f\/lag ${\cal P}^{(d)}$ is made out of
f\/inite-dimensional irreducible representation spaces ${\cal P}_n^{(d)}$ of the
algebra $gl_{d+1}$ taken in realization~\eqref{gln}. It is evident that
{\it any operator made out of generators~\eqref{gln} has finite-dimensional invariant subspace which is
finite-dimensional irreducible representation space.}

\subsubsection{Algebraic properties of the Sutherland model}

It seems evident that the Hamiltonian~\eqref{hAN} has to have a~representation as a~second order polynomial in
generators~\eqref{gln} at $d=N-1$ acting in ${\mathbb R}^{N-1}$,
\begin{gather*}
h_{\rm Suth}=\operatorname{Pol}_2\big({\cal J}_i^-,{{\cal J}_{ij}}^0\big),
\end{gather*}
where the raising generators ${\cal J}_i^+$ are absent. Thus,
$gl(N)$ (or, strictly speaking, its maximal af\/f\/ine subalgebra) is the hidden algebra
of the $N$-body Sutherland model. Hence, $h_{\rm Suth}$ is an element of the universal enveloping algebra ${\cal
U}_{gl(N)}$. The eigenfunctions of the $N$-body Sutherland model are elements of the f\/lag of polynomials ${\cal
P}^{(N-1)}$.
Each subspace ${\cal P}_n^{(N-1)}$ is represented by the Newton polytope (pyramid). It contains $C^{N-1}_{n+N-1}$
eigenfunctions, which is equal to the volume of the Newton polytope. They are orthogonal with respect to $\Psi_0^2$,
see~\eqref{psi_suth}.

The Hamiltonian~\eqref{HAN} is completely-integrable: there exists a~commutative algebra of integrals (including the
Hamiltonian and the momentum of the center-of-mass motion) of dimen\-sion~$N$ which is equal to the dimension of the
conf\/iguration space (for integrals, see Oshima~\cite{Oshima} with explicit forms of those). Each integral ${\cal I}_k$
has a~form polynomial in momentum of degree $k\leq N$. Making gauge rotation with $\Psi_0^2$, separating
center-of-mass motion and changing variable to~\eqref{cms} any integral appears in a~form dif\/ferential operator with
polynomial coef\/f\/icients. Evidently, it preserves the f\/lag of polynomials~\eqref{flag} and can be written as a
non-linear combination of the generators~\eqref{gln} at $d=N-1$ from its af\/f\/ine subalgebra. The explicit formulae of
integrals in~\eqref{gln} are unknown. The spectra of the integral which is a~polynomial in momentum of degree k is
given by a~polynomial in quantum numbers of the degree~$k$. All eigenfunctions of the integrals are common.

Among the generators of the hidden algebra there is the Euler--Cartan operator,
\begin{gather*}
{\cal J}^0_n=\sum_{i=1}^{N-1}\ta_i\frac{\pa}{\pa\ta_i}-n,
\end{gather*}
see~\eqref{gln}, which has zero grading and plays a~role of constant acting as identity operator on a~monomial
in $\ta$. It def\/ines the highest weight vector. This generator allows us to construct the particular
integral~-- $\pi$-integral of zero grading (see~\cite{Turbiner:2012})
\begin{gather}
\label{pics}
i_{\rm par}^{(n)}(\ta)=\prod_{j=0}^n\big({\cal J}^0_n+j\big)
\end{gather}
such that
\begin{gather*}
\big[h_{\rm Suth}(\ta),i_{\rm par}^{(n)}(\ta)\big]: \ {\cal P}^{(N-1)}_{n}\mapsto0.
\end{gather*}
Making the gauge rotation of the $\pi$-integral~\eqref{pics} with $\Psi_{0}^{-1}(\ta)$ given by \eqref{psi_suth} and
changing variables $\ta$ (see~\eqref{cms}) back to the Cartesian coordinates we arrive at the quantum $\pi$-integral,
\begin{gather*}
{\cal I}_{\rm par,Suth}^{(n)}(x)=\left.\Psi_{0}(\ta)i_{\rm par}^{(n)}(\ta)\Psi_{0}^{-1}(\ta)\right|_{\ta\rar x}.
\end{gather*}
It is a~dif\/ferential operator of the $(n+1)$th order.

Under such a~gauge transformation the triangular space of polynomials ${\cal P}^{(N-1)}_{n}$ becomes the space
\begin{gather*}
{\cal V}^{(N-1)}_{n}=\Psi_0{\cal P}^{(N-1)}_{n}.
\end{gather*}
The Hamiltonian ${\cal H}_{\rm Suth}(x)$ commutes with ${\cal I}_{\rm par,Suth}^{(n)}(x)$ over this space
\begin{gather*}
\big[{\cal H}_{\rm Suth}(x),{\cal I}_{\rm par,Suth}^{(n)}(x)\big]: \
{\cal V}^{(N-1)}_{n}\mapsto0.
\end{gather*}
Any eigenfunction $\Psi\in{\cal V}^{(N-1)}_{n}$ is zero mode of the $\pi$-integral ${\cal I}_{\rm par,Suth}^{(n)}(x)$.

\subsection[Case $BC_{N}$]{Case $\boldsymbol{BC_{N}}$}

The $BC_N$-Trigonometric model is def\/ined by the Hamiltonian,
\begin{gather}
{\cal H}_{BC_N}=
-\frac{1}{2}\sum_{i=1}^{N}\frac{\pa^2}{\pa{x_i}^2}+
\frac{g\beta^2}{4}\sum_{i<j}^{N}\left[
\frac{1}{\sin^2\big(\frac{\beta}{2}(x_{i}-x_{j})\big)}+
\frac{1}{\sin^2\big(\frac{\beta}{2}(x_{i}+x_{j})\big)}\right]
\nonumber\\
\phantom{{\cal H}_{BC_N}=}
{}+\frac{g_{2}\beta^2}{2}\sum_{i=1}^{N}
\frac{1}{\sin^2\beta x_{i}}+\frac{g_{3}\beta^2}{8}
\sum_{i=1}^{N}\frac{1}{\sin^2{\frac{\beta x_{i}}{2}}},\label{HBCN}
\end{gather}
where $\beta$, $g$, $g_2$, $g_3$ are parameters. Symmetry: $S_N\oplus(\mathbb{Z}_2)^{\otimes N}\oplus
T$ (permutations $x_i\rar x_j$, ref\/lections $x_i\rar-x_i$,  translation $x_i\rar x_i+2\pi/\beta$). $BC_N$
root space contains roots of the three lengths:~$1$, $\sqrt{2}$, $2$. The $BC_N$ fundamental weights coincide to the $C_N$
fundamental weights.

The ground state function for~\eqref{HBCN} reads
\begin{gather}
\Psi_{0} = \left[
\prod_{i<j}\left|\sin\left(\frac{\beta}{2}(x_{i}\!-\!x_{j})\right)\right|^{\nu}
\left|\sin\left(\frac{\beta}{2}(x_{i}\!+\!x_{j})\right)\right|^{\nu}\right]
\prod_{i=1}^N\big|\sin(\beta x_i)\big|^{\nu_{2}}
\left|\sin\left({\frac{\beta}{2}{x_i}}\right)\right|^{\nu_3}\!,\label{psi_bcn}
\end{gather}
cf.~\eqref{GSES}, where $\nu$, $\nu_2$, $\nu_3$ are found from the relations
\begin{gather*}
g=\nu(\nu-1)>-\frac{1}{4},\qquad g_{2}=\nu_{2}(\nu_{2}-1)>-\frac{1}{4},\qquad
g_{3}=\nu_{3}(\nu_{3}+2\nu_{2}-1)>-\frac{1}{4}.
\end{gather*}
Any eigenfunction has a~form $\Psi_{0}\varphi$, where $\varphi$ is a~polynomial in the $C_N$ fundamental trigonometric
invariants $\ta(\beta)$~\eqref{INV}. Hence, $\Psi_{0}$ plays a~role of multiplicative factor.

The $BC_N$ Hamiltonian~\eqref{HBCN} degenerates to the $B_N$ Hamiltonian at $g_2=0$,
to the $C_N$ Hamiltonian at $g_3=0$ and to the $D_N$ Hamiltonian at $g_2=g_3=0$.
For the $B_N$ Hamiltonian there exist two families of eigenfunctions with multiplicative
factors
\begin{gather*}
\Psi_{0,B_N}^{(1)}=\left[
\prod_{i<j}\left|\sin\left(\frac{\beta}{2}(x_{i}-x_{j})\right)\right|^{\nu}
\left|\sin\left(\frac{\beta}{2}(x_{i}+x_{j})\right)\right|^{\nu}\right]
\left[\left|\sin\left({\frac{\beta}{2}{x_i}}\right)\right|^{\nu_3}\right],
\end{gather*}
and
\begin{gather*}
\Psi_{0,B_N}^{(2)}=\left[
\prod_{i<j}\left|\sin\left(\frac{\beta}{2}(x_{i}-x_{j})\right)\right|^{\nu}
\left|\sin\left(\frac{\beta}{2}(x_{i}+x_{j})\right)\right|^{\nu}\right]
\left[\prod_{i=1}^N\big|\sin(\beta x_i)\big|\right]\!
\left[\left|\sin\left({\frac{\beta}{2}{x_i}}\right)\right|^{\nu_3}\right]\!,
\end{gather*}
respectively.
For the $D_N$ Hamiltonian there exist three families of eigenfunctions with multiplicative
factors
\begin{gather*}
\Psi_{0,D_N}^{(1)}=\left[
\prod_{i<j}\left|\sin\left(\frac{\beta}{2}(x_{i}-x_{j})\right)\right|^{\nu}
\left|\sin\left(\frac{\beta}{2}(x_{i}+x_{j})\right)\right|^{\nu}\right],
\\
\Psi_{0,D_N}^{(2)}=\left[
\prod_{i<j}\left|\sin\left(\frac{\beta}{2}(x_{i}-x_{j})\right)\right|^{\nu}
\left|\sin\left(\frac{\beta}{2}(x_{i}+x_{j})\right)\right|^{\nu}\right]
\left[\prod_{i=1}^N\big|\sin(\beta x_i)\big|\right],
\\
\Psi_{0,D_N}^{(3)}=\left[
\prod_{i<j}\left|\sin\left(\frac{\beta}{2}(x_{i}-x_{j})\right)\right|^{\nu}
\left|\sin\left(\frac{\beta}{2}(x_{i}+x_{j})\right)\right|^{\nu}\right]
\left[
\left|\sin\left({\frac{\beta}{2}{x_i}}\right)\right|\right],
\end{gather*}
respectively.

Let us make a~gauge rotation
\begin{gather*}
h_{{BC}_N}=\frac{1}{\beta^2}(\Psi_{0})^{-1}({\cal H}_{BC_N}-E_0)
\Psi_{0},
\end{gather*}
and then change variables~\cite{Brink:1997}
\begin{gather}
\label{coord_bcn}
(x_1,x_2,\ldots, x_N)\rightarrow\big(\ta_k=\si_k(\cos\beta x)\,\big|\,
{k=1,2,\ldots,N}\big),
\end{gather}
where $\si_{k}$ is the elementary symmetric polynomial, $\ta_0=1$ and $\ta_k=0$
for $k<0$ and $k>N$. It can be checked that $\ta_k$ are $C_N$ trigonometric invariants with
period $\frac{2\pi}{\beta}$. We arrive at~\cite{Brink:1997}
\begin{gather}
\label{hBCN}
{h}_{BC_N}=\sum_{i,j=1}^N{\cal A}_{ij}(\si)\frac{\pa^2}{\pa{\si_i}\pa
{\si_j}}+\sum_{i=1}^N{\cal B}_i(\si)\frac{\pa}{\pa\si_i},
\end{gather}
with coef\/f\/icients
\begin{gather}
{\cal A}_{ij}=
-N{\ta}_{i-1}{\ta}_{j-1}+\sum_{l\ge0}\big[
(i-l){\ta}_{i-l}{\ta}_{j+l}
+(l+j-1){\ta}_{i-l-1}{\ta}_{j+l-1}
\nonumber
\\
\phantom{{\cal A}_{ij}=}
{} -(i-2-l){\ta}_{i-2-l}{\ta}_{j+l}-(l+j+1)
{\ta}_{i-l-1}{\ta}_{j+l+1}\big],\label{Aij}
\\
{\cal B}_i=
\big[1+\nu(2N-i-1)\!+2\nu_2+
\nu_3\big]i{\ta}_{i}\!-\nu_3(i-N-1){\ta}_{i-1}\!
+\nu(N-i+1)(N-i+2){\ta}_{i-2},\!\!\nonumber
\end{gather}
cf.~\eqref{hBC1}.
This is an algebraic form of the $BC_N$ trigonometric Hamiltonian. For polynomial eigenfunctions we f\/ind the
eigenvalues are
\begin{gather*}
\ep_{\{p\}}=\sum_{i=1}^{N}\big[\nu(2N-i-1)+2\nu_2+
\nu_3\big]i p_i+\sum_{i,j=1}^{N}i p_i p_j,
\end{gather*}
cf.~\eqref{EBC1}, hence, the spectrum is quadratic in quantum numbers $p_i=0,1,\ldots$, where $i=1,2,\ldots, N$. The
Hamiltonian $h_{BC_N}$ has inf\/initely many f\/inite-dimensional invariant subspaces of the form ${\cal P}_n^{(N)}$,
see~\eqref{PN}, where $n=0,1,2,\ldots$. They naturally form the f\/lag ${\cal P}^{(N)}$, see~\eqref{flag}. The
Hamiltonian can be immediately rewritten in terms of generators~\eqref{gln} at $d=N$ as a~polynomial of the second
degree,
\begin{gather*}
h_{\rm BC_N}=\operatorname{Pol}_2\big({\cal J}_i^-,{{\cal J}_{ij}}^0\big),
\end{gather*}
where the raising generators ${\cal J}_i^+$ are absent.
Hence, $gl(N+1)$ is the hidden algebra of the $BC_N$ trigonometric model, the same algebra as for the $A_{N}$-rational
model. The eigenfunctions of the~$BC_N$ trigonometric model are elements of the f\/lag of polynomials ${\cal P}^{(N)}$.
Each subspace~${\cal P}_n^{(N)}$ contains $C^{N}_{n+N}$ eigenfunctions (volume of the Newton polytope (pyramid) ${\cal
P}_n^{(N)}$). They are orthogonal with respect to~$\Psi_0^2$, see~\eqref{psi_bcn}.

\begin{figure}[t]\centering
\includegraphics*[width=2.5in,angle=0.0]{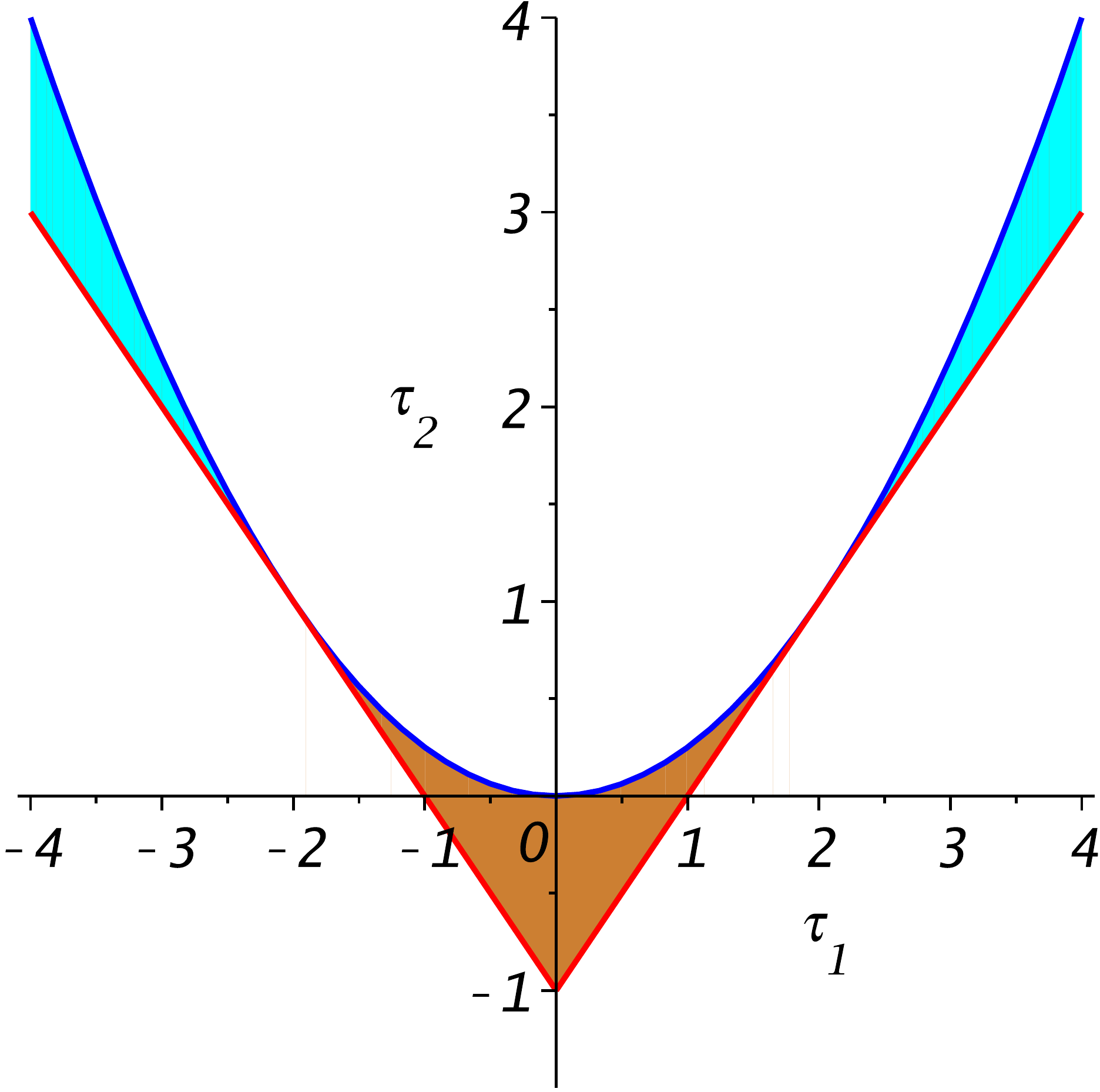}
\caption{An illustration of the conf\/iguration space for $BC_2$ trigonometric model
in $\tau$-variables (light brown area) and for $BC_2$ hyperbolic model (light blue area on the right).}\label{fig2}
\end{figure}

The rational form~\eqref{H-tau-rat} of the $BC_N$ trigonometric Hamiltonian~\eqref{HBCN} can be derived making the
gauge rotation of the algebraic form~\eqref{hBCN} with inverse of the ground state function
in $\ta$-variables, $(\Psi_0(\ta))^{-1}$,
\begin{gather*}
{\cal H}_{BC_N}(\ta)=-\De_g+V_{BC_N}(\ta),
\end{gather*}
where $\De_g$ is the Laplace--Beltrami operator with a~metric $g^{ij}(\ta)={\cal A}_{ij}$ (see~\eqref{Aij})
and $V_{BC_N}(\ta)$ is a~potential. The explicit expression for $V_{BC_1}(\ta)$
is presented in~\eqref{HBC1-tau} while the ground state eigenfunction $\Psi_0^{(BC_1)}(\ta)$
is given by~\eqref{psi0-BC1T}. The conf\/iguration space in $\tau$ coordinate is the
interval, $\tau\in[-1,1]$ (trigonometric case) or half-line, $\tau\in[1,\infty)$ (hyperbolic case). As for $BC_2$
case,
\begin{gather}
\label{HBC2-tau}
V_{BC_2}(\ta)=g\frac{1-\ta_2}{\ta_1^2-4\ta_2}+
\frac{g_2}{4}\frac{2-\ta_1}{1+\ta_1+\ta_2}+
\frac{1}{4}\frac{2(g_2+g_3)+g_2\ta_1-g_3\ta_2}{1-\ta_1+\ta_2},
\end{gather}
and
\begin{gather*}
\Psi_0^{(BC_2)}(\ta)=\big(\ta_1^2-4\ta_2\big)^{\frac{\nu}{2}}
(1+\ta_1+\ta_2)^{\frac{\nu_2}{2}}(1-\ta_1+\ta_2)^{\frac{\nu_2+\nu_3}{2}},
\end{gather*}
and the conf\/iguration space is illustrated by Fig.~\ref{fig2}.

As for $BC_3$
\begin{gather}
V_{BC_3}(\ta)=g\frac{\ta_1^4-\ta_1^3\ta_3-6\ta_1^2\ta_2+
9
\ta_1
\ta_2
\ta_3
+
9\ta_2^2-\ta_2^3-27\ta_3^2}{\ta_1^2\ta_2^2-4\ta_1^3\ta_3-4\ta_2^2-27\ta_3^2+18\ta_1\ta_2\ta_3}\nonumber\\
\hphantom{V_{BC_3}(\ta)=}{} +
\frac{g_2}{2}
\frac{3+2\ta_1+\ta_2}{1+\ta_1+\ta_2+\ta_3}
+
\frac{g_2+4g_3}{4}\frac{3-2\ta_1+\ta_2}{1-\ta_1+\ta_2-\ta_3},\label{HBC3-tau}
\end{gather}
and
\begin{gather*}
\Psi_0^{(BC_3)}(\ta)
=\big(\ta_1^2\ta_2^2-4\ta_1^3\ta_3-4\ta_2^2-27\ta_3^2+18\ta_1\ta_2\ta_3\big)^{\frac{\nu}{2}} \\
\hphantom{\Psi_0^{(BC_3)}(\ta)=}{}
\times
(1+\ta_1+\ta_2+\ta_3)^{\frac{\nu_2}{2}}(1-\ta_1+\ta_2-\ta_3)^{\frac{\nu_2+\nu_3}{2}}.
\end{gather*}

The Hamiltonian~\eqref{HBCN} is completely-integrable: there exists a~commutative algebra of integrals (including the
Hamiltonian) of dimension $N$ which is equal to the dimension of the conf\/iguration space (for integrals, see
Oshima~\cite{Oshima} with explicit forms of those). Each integ\-ral~${\cal I}_k$ has a~form polynomial in momentum of
degree $2k\leq2N$. Making gauge rotation
with~$\Psi_0^2$ and changing variable to~\eqref{cms} any integral appears in a~form dif\/ferential operator with
polynomial coef\/f\/icients. Evidently, it preserves the f\/lag of polynomials~\eqref{flag} and can be written as a
non-linear combination of the generators~\eqref{gln} at $d=N$ from its af\/f\/ine subalgebra. The explicit formulae of
integrals in generators~\eqref{gln} are unknown. The spectra of the integral which is a~polynomial in momentum of
degree~$2k$ is given by a~polynomial in quantum numbers of the degree~$2k$. All eigenfunctions of the integrals are
common.

It is evident that for the $BC_N$ trigonometric model there exists a~particular integral~-- $\pi$-integral of zero
grading (see~\cite{Turbiner:2012})
\begin{gather*}
i_{\rm par}^{(n)}(\ta)=\prod_{j=0}^n\big({\cal J}^0_n+j\big)
\end{gather*}
(cf.~\eqref{pics}), such that
\begin{gather*}
\big[h_{BC_N}(\ta),i_{\rm par}^{(n)}(\ta)\big]: \ {\cal P}^{(N)}_{n}\mapsto0.
\end{gather*}
Making the gauge rotation of the $\pi$-integral~\eqref{pics} with $\Psi_{0}^{-1}(\ta)$ given by \eqref{psi_bcn} and
changing variables $\ta$ (see~\eqref{coord_bcn}) back to the Cartesian coordinates we arrive at the
quantum $\pi$-integral,
\begin{gather*}
{\cal I}_{{\rm par},BC_N}^{(n)}(x)=\Psi_{0}(\ta)i_{\rm par}^{(n)}(\ta)\Psi_{0}^{-1}(\ta)\big|_{\ta\rar x}.
\end{gather*}
It is a~dif\/ferential operator of the $(n+1)$th order.

Under such a~gauge transformation the triangular space of polynomials ${\cal P}^{(N)}_{n}$ becomes the space
\begin{gather*}
{\cal V}^{(N)}_{n}=\Psi_0{\cal P}^{(N)}_{n}.
\end{gather*}
The Hamiltonian ${\cal H}_{BC_N}(x)$ commutes with ${\cal I}_{{\rm par},BC_N}^{(n)}(x)$ over this space{\samepage
\begin{gather*}
\big[{\cal H}_{BC_N}(x),{\cal I}_{{\rm par},BC_N}^{(n)}(x)\big]: \
{\cal V}^{(N)}_{n}\mapsto0.
\end{gather*}
Any eigenfunction $\Psi\in{\cal V}^{(N)}_{n}$ is zero mode of the $\pi$-integral ${\cal I}_{{\rm par},BC_N}^{(n)}(x)$.}

Now we are in a~position to draw an intermediate conclusion about $A_N$ and $BC_N$ trigonometric models.
\begin{itemize}\itemsep=0pt
\item Both $A_N$- and $BC_N$-trigonometric (and rational) models possess
{\it algebraic} forms associated with preservation of the {\it same}
f\/lag of polynomials ${\cal P}^{(N)}$.
The f\/lag is invariant with respect to linear transformations in space of orbits
$\ta\mapsto\ta+A$. It preserves the algebraic form of Hamiltonian.

\item
Their Hamiltonians (as well as higher integrals) can be written
in the Lie-algebraic form
\begin{gather*}
h=\operatorname{Pol}_2({\cal J}(b\subset gl_{N+1})),
\end{gather*}
where $\operatorname{Pol}_2$ is a~polynomial of 2nd degree in generators
${\cal J}$ of the maximal af\/f\/ine subalgebra of the algebra $b$ of the algebra
$gl_{N+1}$ in realization~\eqref{gln}.
Hence, $gl_{N+1}$ is their {\it hidden algebra}. From this viewpoint all
four models are dif\/ferent faces of a~{\it single} model.

\item {\it\sloppy  Supersymmetric $A_N$- and $BC_N$-rational $($and trigonometric$)$
models possess {\it algebraic} forms, preserve the {\it same} flag
of $($super$)$polynomials and their {\it hidden algebra} is the superalgebra
$gl(N+1|N)$ $($see~{\rm \cite{Brink:1997})}.

}
\end{itemize}

In a~connection to f\/lags of polynomials we introduce a~notion
`{\it characteristic vector}'.
Let us consider a~f\/lag made out of ``triangular'' linear space of
polynomials
\begin{gather*}
{\cal P}^{(d)}_{n,\vec f}=\langle x_1^{p_1}
x_2^{p_2}\cdots x_d^{p_d}\, |\, 0\leq f_1p_1+f_2p_2+\cdots+
f_d p_d\leq n\rangle,
\end{gather*}
where the ``grades'' $f$'s are positive integer numbers and $n=0,1,2,\ldots$.
In lattice space ${\cal P}^{(d)}_{n,\vec f}$ def\/ines a~Newton pyramid.

\begin{definition}
Characteristic vector is a~vector with components
$f_i$:
\begin{gather*}
\vec f=(f_1,f_2,\ldots, f_d).
\end{gather*}
From geometrical point of view $\vec f$ is normal vector to the base of the Newton pyramid. The characteristic vector
for f\/lag ${\cal P}^{(d)}$ is
\begin{gather*}
\vec f_0=\underbrace{(1,1,\ldots, 1)}_{d}.
\end{gather*}
\end{definition}

\subsection[Case $G_2$]{Case $\boldsymbol{G_2}$}

Take the Hamiltonian
\begin{gather*}
{\cal H}_{G_2}=
-\frac{1}{2}\sum_{k=1}^{3}\frac{\pa^{2}}{\pa x_{k}^{2}}
+\frac{g\beta^2}{4}\sum_{k<l}^{3}\frac{1}{\sin^{2}
(\frac{\beta}{2}(x_{k}-x_{l}))}
+\frac{g_1\beta^2}{4}\sum_{k<lk,\, l\neq m}^{3}
\frac{1}{\sin^{2}(\frac{\beta}{2}(x_{k}+x_{l}-2x_{m}))},
\end{gather*}
where $g$, $g_1$ and $\beta$ are parameters.
It describes a~trigonometric generalization of the rational Wolfes model of three-body interacting system or, in the
Hamiltonian reduction nomenclature, the $G_2$-trigonometric model~\cite{OP}.
The symmetry of the model is dihedral group $D_6\oplus T$. The ground state function is
\begin{gather*}
\Psi_{0}=\prod_{i<j}^3\left|\sin\frac{\beta}{2}(x_{i}-x_{j})\right|^{\nu}
\prod_{k<lk,\, l\neq m}^3\left|\sin\frac{\beta}
{2}(x_{i}+x_{j}-2x_{k})\right|^{\mu}
\end{gather*}
with $\nu,\mu>-\frac{1}{2}$ as solutions of
\begin{gather*}
g=\nu(\nu-1)>-\frac{1}{4},\qquad g_1=3\mu(\mu-1)>
-\frac{3}{4}.
\end{gather*}

Making the gauge rotation
\begin{gather*}
h_{{G}_2}=(\Psi_{0})^{-1}({\cal H}_{G_2}-E)\Psi_{0},
\end{gather*}
and changing variables~\cite{Rosenbaum:1998}
\begin{gather*}
Y=\sum x_i, \qquad y_i=x_i-\frac{1}{3}Y,\quad i=1,2,3,
\qquad
(x_1,x_2,x_3)\rightarrow\big(Y,\ta_1,\ta_2\big),
\end{gather*}
where
\begin{gather*}
\ta_1=2[\cos(\beta(y_1-y_2))+
\cos(\beta(2y_1+y_2))+\cos(\beta(y_1+2y_2))],
\\
\ta_2=2[\cos(3\beta y_1)+ \cos(3\beta y_2)+\cos(3\beta(y_1+y_2))]
\end{gather*}
are $G_2$ trigonometric invariants, and separating the center-of-mass coordinate we arrive at~\cite{Rosenbaum:1998}
\begin{gather}
h_{G_2}=-\left(4+\ta_1+\frac{\tau_2}{3}-\frac{\ta_1^2}{3}\right)
\pa_{\tau_1\tau_1}^2+\big(12+4\ta_2+\tau_1\ta_2-2\ta_1^2\big)
\pa_{\tau_1\tau_2}^2\nonumber\\
\hphantom{h_{\rm G_2}=}{}
+\big(9\ta_1+3\ta_2+3\ta_1\ta_2+\ta_2^2-\ta_1^3\big)
\pa_{\tau_2\ta_2}^2
+\left[2\nu+\frac{1+3\mu+2\nu}{3}\ta_1\right]\pa_{\ta_1}\nonumber\\
\hphantom{h_{\rm G_2}=}{}
+\big[6\mu+(1+2\mu+\nu)\tau_2+2\nu\ta_1\big]
\pa_{\ta_2},\label{G2-alg}
\end{gather}
which is the algebraic form of the $G_2$ trigonometric Hamiltonian. The eigenvalues of $h_{G_2}$ are
\begin{gather*}
\ep_{\{p\}}=\frac{p_1^2}{3}+p_1p_2+p_2^2+(\mu+\nu)p_1+(2\mu+\nu)p_2
\end{gather*}
quadratic in quantum numbers $p_1,p_2=0,1,2,\ldots$.

The Hamiltonian $h_{G_2}$ has inf\/initely many f\/inite-dimensional invariant subspaces
\begin{gather}
\label{P12}
{\cal P}_{n,(1,2)}^{(2)}=\langle{\ta_{1}}^{p_1}
{\ta_{2}}^{p_2}\,\vert\, 0\le p_1+2p_2\le n\rangle,\qquad n=0,1,2,\ldots,
\end{gather}
hence the f\/lag ${\cal P}^{(2)}_{(1,2)}$ with the characteristic vector $\vec f=
(1,2)$ is preserved by $h_{G_2}$. The eigenfunctions of $h_{G_2}$ are
are elements of the f\/lag ${\cal P}^{(2)}_{(1,2)}$.
Each space $({\cal P}_{n,(1,2)}^{(2)}\ominus{\cal P}_{n-1,(1,2)}^{(2)})$ contains ${\sim}n$ eigenfunctions
which is equal to length of the Newton line
${\cal L}_n=\langle{\ta_{1}}^{p_1}{\ta_{2}}^{p_2}\vert p_1+2p_2=n\rangle$.

A natural question to ask whether does an algebra of dif\/ferential operators
exist for which ${\cal P}_{n,(1,2)}^{(2)}$ is the space of (irreducible)
representation. We call this algebra $g^{(2)}$~\cite{Rosenbaum:1998}.

\subsection[Algebra $g^{(2)}$]{Algebra $\boldsymbol{g^{(2)}}$}

\begin{figure}[t]\centering

\includegraphics{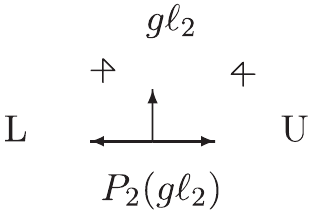}

\caption{Triangular diagram relating the subalgebras $L$,
$U$ and $g\ell_2$. ${P}_2{(g\ell_2)}$ is a~polynomial of the 2nd degree in $g\ell_2$ generators. It is a
generalization of the Gauss
decomposition for semi-simple algebras.}\label{Fig3}
\end{figure}

Let us consider the Lie algebra spanned by seven generators
\begin{gather}
J^1=\pa_t,\qquad
J^2_n=t\pa_t-\frac{n}{3}, \qquad
J^3_n=2u\pa_u-\frac{n}{3},\qquad J^4_n=t^2\pa_t+2t u\pa_u-n t,
\nonumber\\
R_{i}=t^{i}\pa_u,\quad i=0,1,2,\qquad{\cal R}^{(2)}\equiv(R_{0},R_{1},R_{2}).\label{gl2r}
\end{gather}
It is non-semi-simple algebra $gl(2,{\mathbb R})\ltimes{\cal R}^{(2)}$
(S.~Lie~\cite[p.~767--773]{Lie} at $n=0$ and A.~Gonz\'alez-Lop\'ez et al.~\cite{glko} at $n\neq0$ (case 24)). If the
parameter~$n$ in~\eqref{gl2r} is a~non-negative integer, it has~\eqref{P12}
\begin{gather*}
{\cal P}_n^{(2)}=\big(t^{p}u^{q}\,|\,0\leq(p+2q)\leq n\big),
\end{gather*}
as common (reducible) invariant subspace. By adding three operators
\begin{gather}
\label{gl2t}
T_0=u\pa_{t}^2,\qquad T_1=u\pa_{t}J_0^{(n)},\qquad T_2=
u J_0^{(n)}\big(J_0^{(n)}+1\big)=u J_0^{(n)}J_0^{(n-1)},
\end{gather}
where
\begin{gather*}
J_0^{(n)}=t\pa_t+2u\pa_u-n,
\end{gather*}
to $gl(2,{\mathbb R})\ltimes{\cal R}^{(2)}$ (see~\eqref{gl2r}),  the action on ${\cal P}_{n,(1,2)}^{(2)}$ gets
irreducible.
Multiple commutators of~$J^4_n$ with~$T_0^{(2)}$ generate new operators acting on ${\cal P}_{n,(1,2)}^{(2)}$,
\begin{gather*}
T_i\equiv\underbrace{[J^4,[J^4,[\ldots J^4,T_0]\ldots]}_i=
u\pa_{t}^{2-i}J_0^{(n)}\big(J_0^{(n)}+1\big)\cdots(J_0^{(n)}+i-1)\\
\hphantom{T_i\equiv\underbrace{[J^4,[J^4,[\ldots J^4,T_0]\ldots]}_i}{}
=
u\pa_{t}^{2-i}\prod_{j=0}^{i-1}J_0^{(n-j)},\qquad i=0,1,2,
\end{gather*}
all of them are dif\/ferential operators of degree $2$. These new generators have a~property of nilpotency,
\begin{gather*}
T_i=0,\qquad i>2,
\end{gather*}
and commutativity:
\begin{gather*}
[T_i,T_j]=0,\qquad i,j=0,1,2,\quad{\cal U}^{(2)}\equiv(T_{0},T_{1},T_{2}).
\end{gather*}
The generators~\eqref{gl2r} plus~\eqref{gl2t} span a~linear space with a~property of
decomposition:  $g^{(2)}\doteq{\cal R}^{(2)}\rtimes(gl_2\oplus J_0)\ltimes{\cal U}^{(2)}$ (see Fig.~\ref{Fig3}).

It is worth mentioning a~property of conjugation ${\cal R}^{(2)}\Leftrightarrow{\cal T}^{(2)}$:
\begin{gather*}
\pa_{\ta_2}\leftrightarrow{\ta_2}J_0^{(n)}\big(J_0^{(n)}+1\big),
\qquad
{\ta_1}\pa_{\ta_2}\leftrightarrow{\ta_2}\pa_{\ta_1}J_0^{(n)},
\qquad
{\ta_1}^{2}\pa_{\ta_2}\leftrightarrow{\ta_2}\pa_{\ta_1}^{2}.
\end{gather*}
where $J_0^{(n)}={\ta_1}\pa_{\ta_1}+2{\ta_2}\pa_{\ta_2}-n$.

Eventually, {\it infinite-dimensional, eleven-generated algebra $($by~\eqref{gl2r} and $J_0$ plus \eqref{gl2t}, so that
the eight generators are the $1$st order and three generators are of the $2$nd order differential operators$)$} occurs. The
Hamiltonian $h_{G_2}$ can be rewritten in terms of the genera\-tors~\eqref{gl2r},~\eqref{gl2t} with the absence of
the highest weight generator~$J^4_n$,
\begin{gather*}
h_{G_2}=-\big(4J^1+J^2-2J^3-12R_0+2R_2\big)J^1+\frac{1}{6}\big(2J^2+3J^3\big)J^2
+\left(J^3+\frac{3}{2}R_1\right)J^3
\\
\hphantom{h_{G_2}=}{} +(9R_0 -R_2)R_1-\frac{1}{3}T_0
+2\nu J^1+\frac{3\mu\!+\!2\nu}{3}J^2+\frac{2\mu\!+\!\nu\!-\!1}{2}J^3+6\mu R_{0}+\left(2\nu-\frac{3}{2}\right)R_{1}
\end{gather*}
(see~\cite{Rosenbaum:1998}), where $J^{2,3}\equiv J^{2,3}_0$. Hence, $gl(2,{\mathbb R})\ltimes{\cal R}^{(2)}$ is the
hidden algebra of the~$G_2$ trigonometric model.

The $G_2$ trigonometric Hamiltonian admits the integral in a~form of the 6th order
dif\/ferential operator~\cite{Oshima}. After gauge rotation with $\Psi_0$ in variables $\ta_{1,2}$
the integral has to take the algebraic form which is not known explicitly. This integral preserves the same f\/lag ${\cal
P}^{(2)}_{(1,2)}$ as the Hamiltonian~\eqref{G2-alg}. It can be rewritten in term of generators of the
algebra $g^{(2)}$. In addition to it, there exists $\pi$-integral
of zero grading (see~\cite{Turbiner:2012})
\begin{gather*}
i_{\rm par}^{(n)}(\ta)=\prod_{j=0}^n\big(J_0^{(n)}+j\big)=
\prod_{j=0}^n J_0^{(n-j)}
\end{gather*}
(cf.~\eqref{pics}), such that
\begin{gather*}
\big[h_{G_2}(\ta),i_{\rm par}^{(n)}(\ta)\big]: \ {\cal P}_{n,(1,2)}^{(2)}\mapsto0.
\end{gather*}
Making the gauge rotation of the $\pi$-integral~\eqref{pics} with $\Psi_{0}^{-1}(\ta)$ given by \eqref{psi_bcn} and
changing variables $\ta$ (see~\eqref{coord_bcn}) back to the Cartesian coordinates we arrive at the
quantum $\pi$-integral,
\begin{gather*}
{\cal I}_{{\rm par},G_2}^{(n)}(x)=\Psi_{0}(\ta)i_{\rm par}^{(n)}(\ta)\Psi_{0}^{-1}(\ta)\big|_{\ta\rar x}.
\end{gather*}
It is a~dif\/ferential operator of the $(n+1)$th order.

Under such a~gauge transformation the triangular space of polynomials ${\cal P}_{n,(1,2)}^{(2)}$ becomes the space
\begin{gather*}
{\cal V}^{(N)}_{n}=\Psi_0{\cal P}_{n,(1,2)}^{(2)}.
\end{gather*}
The Hamiltonian ${\cal H}_{G_2}(x)$ commutes with ${\cal I}_{{\rm par},G_2}^{(n)}(x)$ over this space
\begin{gather*}
\big[{\cal H}_{G_2}(x),{\cal I}_{{\rm par},G_2}^{(n)}(x)\big]: \
{\cal V}^{(N)}_{n}\mapsto0.
\end{gather*}
Any eigenfunction $\Psi\in{\cal V}^{(N)}_{n}$ is zero mode of the $\pi$-integral ${\cal I}_{{\rm par},G_2}^{(n)}(x)$.

Summarizing let us mention that in addition to the f\/lag ${\cal P}^{(2)}_{(1,2)}$ the $G_2$ trigonometric Hamiltonian
preserves two more f\/lags:
${\cal P}_{(3,5)}$ and ${\cal P}_{(5,9)}$,
where their characteristic vectors $(3,5)$ and~$(5,9)$ coincide to the Weyl vector and co-vector, respectively.

\subsection[Cases $F_4$ and $E_{6,7}$]{Cases $\boldsymbol{F_4}$ and $\boldsymbol{E_{6,7}}$}

These three cases are described in some details in~\cite{BLT:2008, VGT:2009} and in \cite[p.~1416]{BTLG:2011},
respectively.

\subsection[Case $E_{8}$ (in brief)]{Case $\boldsymbol{E_{8}}$ (in brief)}

In this Section a~brief description of $E_{8}$ trigonometric case is given, all details can be found
in~\cite{BTLG:2011}.

The $E_{8}$ trigonometric Hamiltonian has a~form~\eqref{HES},
\begin{gather}
\mathcal{H}_{E_8}\left(\frac{\beta}{2}\right)=-\frac{1}{2}\Delta^{(8)}+\frac{g\beta^2}{4}\sum_{j<i=1}^{8}
\left[\frac{1}{\sin^{2}{\frac{\beta}{2}(x_i+x_j)}}+\frac{1}{\sin^{2}
{\frac{\beta}{2}(x_i-x_j)}}\right]
\nonumber\\
\hphantom{\mathcal{H}_{E_8}\left(\frac{\beta}{2}\right)=}{}
+\frac{g\beta^2}{4}\sum_{\{\nu_j\}}\frac{1}{\Big[\sin^{2}
\frac{\beta}{4}\Big({x_8+\sum\limits_{j=1}^7
(-1)^{\nu_j}x_j}\Big)\Big]} , \label{H_E8}
\end{gather}
and it acts in ${\mathbb R}^8$. The second summation being one over septuples $\{\nu_j\}$ where each $\nu_j=0,1$
and $\sum\limits_{j=1}^{7}\nu_j$ is even. Here $g=\nu(\nu-1)>-1/4$ is the coupling constant and $\beta$ is a
parameter. The conf\/iguration space is the principal $E_8$ Weyl alcove.
Symmetry of the~$E_{8}$ trigonometric model is given by the af\/f\/ine ${\rm E_8}$ Weyl group of the order 696\,729\,600.
The ground state function $\Psi_{0}$ is given by~\eqref{GSES}.
Making a~gauge rotation of the Hamiltonian
\begin{gather*}
h_{E_8}=\frac{1}{\beta^2}(\Psi_{0})^{-1}(\mathcal{H}_{E_8}-E_{0})\Psi_{0},
\end{gather*}
where $E_0=310\beta^2\nu^2$ is the ground state energy,
and introducing the new variables $\ta_{1,\ldots,8}(\beta)$, which are the fundamental trigonometric invariants with
respect to the ${\rm E_8}$ Weyl group, we arrive at the $E_{8}$ trigonometric Hamiltonian in the algebraic form
\begin{gather}
\label{he8}
h_{E_8}=\sum_{i,j=1}^{4}A_{ij}(\ta)\frac{\pa^2}{\pa\tau_i\pa\tau_j}+
\sum_{j=1}^{4}B_j(\ta,\nu)\frac{\pa}{\pa\tau_j},
\end{gather}
where $A_{ij}(\ta)$, $B_j(\ta;\nu)$ are polynomials in $\ta$ with integer coef\/f\/icients and
$B_j(\ta;\nu)$ depend on $\nu$ linearly (see~\cite[Appendix~A]{BTLG:2011}).

It is easy to check that the algebraic operator $h_{E_8}$ has inf\/initely-many f\/inite-dimensional invariant subspaces
\begin{gather*}
\mathcal{P}_{n}^{(2,2,3,3,4,4,5,6)}=
\langle
\ta_{1}^{n_{1}}\ta_{2}^{n_{2}}\ta_{3}^{n_{3}}\ta_{4}^{n_{4}}
\ta_{5}^{n_{5}}\ta_{6}^{n_{6}}\ta_{7}^{n_{7}}\ta_{7}^{n_{8}}\, | \\
\hphantom{\mathcal{P}_{n}^{(2,2,3,3,4,4,5,6)}= \langle}{} 0\leq
2n_{1}+2n_{2}+3n_{3}+3n_{4}+4n_{5}+4n_{6}+5n_{7}+6n_{8}\leq n\rangle,
\qquad n\in\mathbb{N},
\end{gather*}
all of them have with the same characteristic vector
$\vec f=(2,2,3,3,4,4,5,6)$, they form the inf\/inite f\/lag. The spectrum of the Hamiltonian $h_{E_8}$~\eqref{he8} is
quadratic in quantum numbers~\cite{BTLG:2011, Sasaki:2000}.

Eigenfunctions $\phi_{n,\{p\}}$ of $h_{E_8}$ are elements of $\mathcal{P}_n^{(2,2,3,3,4,4,5,6)}$. The number of
eigenfunctions in $\mathcal{P}_n^{{(2,2,3,3,4,4,5,6)}}$ is equal to the dimension
of $\mathcal{P}_n^{{(2,2,3,3,4,4,5,6)}}$.

\begin{figure}[t]\centering
\includegraphics{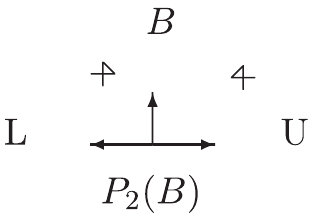}

\caption{Triangular diagram relating the subalgebras $L$,
$U$ and $B$. ${P}_p{(B)}$ is a~polynomial of the $p$th degree in $B$ generators. It is a~generalization of the Gauss
decomposition for semi-simple algebras.}\label{Fig4}
\end{figure}

The space $\mathcal{P}^{{(2,2,3,3,4,4,5,6)}}_{n}$ is a~f\/inite-dimensional representation space of a~Lie algebra of
dif\/ferential operators which we call the $e^{(8)}$ algebra~\cite{G-T:2012}. It is inf\/inite-dimensional but f\/initely
generated algebra of dif\/ferential operators, with 968 generating elements in a~form of dif\/ferential operators of the
orders
$1^{\text{st}}$
(54), $2^{\text{nd}}$ (24), $3^{\text{rd}}$ (18), $4^{\text{rd}}$ (18), $5^{\text{rd}}$ (28), $6^{\text{rd}}$ (5) plus
one of zeroth order (constant). They span 100 + 100 Abelian (conjugated) subalgebras of lowering and raising
generators\footnote{It implies that these commutative subalgebras can be divided into pairs.
In every pair the elements of dif\/ferent subalgebras are related via a~certain operation of conjugation similar to one
described for $g^{(2)}$ on p.~18.} $L$ and $U$ and one algebra $B$ of the Cartan type of dimension 15 plus one central element. Among
the generators of $B$ there is the Euler--Cartan operator
\begin{gather}
\label{J0-e8}
J_0^{(n)}=2\ta_1\pa_{\ta_1}+2\ta_2\pa_{\ta_2}+
3\ta_3\pa_{\ta_3}+3\ta_4\pa_{\ta_4}+
4\ta_5\pa_{\ta_5}+4\ta_6\pa_{\ta_6}+
5\ta_7\pa_{\ta_7}+6\ta_8\pa_{\ta_8}-n.
\end{gather}
Taking the algebra $B$ and a~pair of conjugated Abelian algebras one can show that the commutation relations lead to
the diagram of Fig.~\ref{Fig4}. Depending on what pair $L$, $U$ the degree $p$ takes the following
values: 2, 3, 4, 5, 6, 7, 8, 9, 10.

The $E_{8}$ trigonometric model is completely-integrable~-- there exist seven algebraically independent mutually
commuting dif\/ferential operators of f\/inite order that commute with the Hamiltonian~\eqref{H_E8}~\cite{Sasaki:2000, OP}.
We are not aware on the existence of their explicit forms. It seems evident that any of these integrals after the gauge
rotation with the ground state function $\Psi_0$ the space of orbits should take an algebraic form of a~dif\/ferential
operator with polynomial coef\/f\/icient functions. Any integral as well as the Hamiltonian is an element of the
algebra~$e^{(8)}$. In addition to ``global'' integrals, there exists $\pi$-integral of zero
grading (see~\cite{Turbiner:2012})
\begin{gather*}
i_{\rm par}^{(n)}(\ta)=\prod_{j=0}^n\big(J_0^{(n)}+j\big)=
\prod_{j=0}^n J_0^{(n-j)},
\end{gather*}
where $J_0^{(n)}$ is given by~\eqref{J0-e8} (cf.~\eqref{pics}) such that
\begin{gather*}
\big[h_{E_8}(\ta),i_{\rm par}^{(n)}(\ta)\big]: \ \mathcal{P}_n^{(2,2,3,3,4,4,5,6)}\mapsto0.
\end{gather*}

It is worth mentioning that the operator~\eqref{he8} has a~certain property of degeneracy: it also preserves the
inf\/inite f\/lag of the spaces of polynomials with the characteristic vector $\vec f=(29,46,57,68,84,91,110,135)$. This
vector coincides to the~$E_8$
Weyl (co)vector. Hence, the eigenfunctions of $h_{E_8}(\ta)$ are the elements of this f\/lag as well. It implies the
existence of another $\pi$-integral $\tilde i_{\rm par}^{(n)}(\ta)$ with $J_0^{(n)}$ given by
\begin{gather*}
J_0^{(n)}=29\ta_1\pa_{\ta_1}\!+46\ta_2\pa_{\ta_2}\!+
57\ta_3\pa_{\ta_3}\!+68\ta_4\pa_{\ta_4}\!+
84\ta_5\pa_{\ta_5}\!+91\ta_6\pa_{\ta_6}\!+
110\ta_7\pa_{\ta_7}\!+135\ta_8\pa_{\ta_8}\!-n,
\end{gather*}
such that
\begin{gather*}
\big[h_{E_8}(\ta),\tilde i_{\rm par}^{(n)}(\ta)\big]: \ \mathcal{P}_n^{(29,46,57,68,84,91,110,135)}\mapsto0.
\end{gather*}

\section{Conclusions}

\begin{itemize}\itemsep=0pt
\item
For trigonometric Hamiltonians for all classical $A_N$, $BC_N$, $B_N$, $C_N$, $D_N$ and for exceptional root
spaces $G_2$, $F_4$, $E_{6,7,8}$, similar to the rational Hamiltonians including
non-crystallographic $H_{3,4}$, $I_{2}(k)$ (see~\cite{Turbiner:2011}), there exists an algebraic form after gauging
away the ground state eigenfunction, and changing variables from Cartesian to fundamental trigonometric Weyl
invariants (see~\cite{Boreskov:2001, BLT:2008, BTLG:2011, Brink:1997, VGT:2009, Rosenbaum:1998, RT:1995}). Their
eigenfunctions are polynomials in these variables. They are orthogonal with respect to the squared ground state
eigenfunction.

Coef\/f\/icient functions in front of the second derivatives of these gauge-rotated Hamiltonians which are polynomials in
fundamental trigonometric Weyl invariants def\/ine a~metric ${\cal A}$ of f\/lat space in the space of orbits. We will call
this metric the {\it V.I.~Arnold metric}, he was the f\/irst to calculate a~similar metric in the case of polynomial Weyl
invariants.
This metric has a~property that in the Laplace--Beltrami operator the coef\/f\/icient functions in front of the f\/irst
derivatives are polynomials in fundamental trigonometric invariants. This property is similar to one which occurs in
the case of rational models. The (rational) Arnold metric for the space of orbits parameterized by polynomial Weyl
invariants can be considered as an appropriate degeneration of the (trigonometric) Arnold metric for the space of
orbits parameterized by fundamental trigonometric Weyl invariants.

\item
Any trigonometric Hamiltonian is characterized by a~hidden algebra. These hidden algebras are $U_{gl(N+1)}$ for the
case of classical $A_N$, $BC_N$, $B_N$, $C_N$, $D_N$ and {\it new} inf\/inite-dimensional but f\/inite-generated algebras
of dif\/ferential operators for all other cases. All these algebras have f\/inite-dimensional invariant subspace(s) in
polynomials.
Rational Hamiltonians are characterized by the same hidden algebra with a~single exception of the $E_8$ case.

\item
The generating elements of any such hidden algebra can be grouped into an even number of (conjugated) Abelian
algebras $L_i$, $U_i$ and one Lie algebra $B$. They obey a~(generalized) Gauss decomposition rule (see
Fig.~\ref{Fig5}). A study and a~description of all these algebras is in progress and will be given elsewhere.

\begin{figure}[t]\centering
\includegraphics{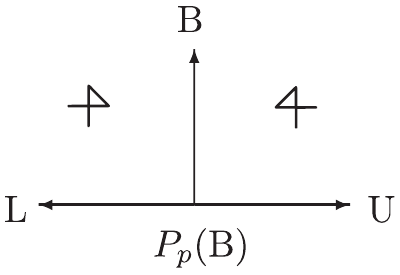}
\caption{Triangular diagram relating the subalgebras $L$,
$U$ and $B$. ${P}_p{(B)}$ is a~polynomial of the $p$th degree in~$B$ generators. It is a~generalization of the Gauss
decomposition for semi-simple algebras where $p=1$.}\label{Fig5}
\end{figure}
\begin{table}[t]\centering
\caption{Minimal characteristic vectors for rational (non)crystallographic and
trigonometric crystallographic systems (see~\cite{BTLG:2011}). For latter case
the Weyl vector and co-vector as possible characteristic vectors occur.
Characteristic vectors for $H_3$, $H_4$, $I_2(k)$ are from~\cite{H3, H4, TTW:2009}, respectively.}\label{table1}

\vspace{1mm}\small

\begin{tabular}{|@{\,\,}c@{\,\,}|@{\,\,}c@{\,\,}|c@{\,\,}|@{\,\,}c@{\,\,}|@{\,\,}c@{\,\,}|}
\hline
  \raisebox{-2ex}[0pt][0pt]{Model}   & \raisebox{-2ex}[0pt][0pt]{Rational} & \multicolumn{3}{@{\,\,}c@{\,\,}|}{Trigonometric} \\ \cline{3-5}
& &   minimal &  integer Weyl&  integer co-Weyl \\
\hline \hline
$A_N$\tsep{2pt} & $\underbrace{(1,1,\ldots,1)}_{N}$\bsep{6pt} &
$\underbrace{(1,1,\ldots,1)}_{N}$ && \\
\hline
 $BC_N$\tsep{2pt} & $\underbrace{(1,1,\ldots,1)}_{N}$ &
$\underbrace{(1,1,\ldots, 1)}_{N}$ && \\
\hline
$G_2$ & (1,2) & (1,2) & (3,5) & $(5,9)$\\
\hline
$F_4$ & (1,2,2,3) & (1,2,2,3) & (8,11,15,21)& (11,16,21,30)\\
\hline
$E_6$ & (1,1,2,2,2,3) & (1,1,2,2,2,3) & (8,8,11,15,15,21)& (8,8,11,15,15,21)\\
\hline
$E_7$ & (1,2,2,2,3,3,4) & (1,2,2,2,3,3,4) &
$(27,34,49,52,66,75,96)$&$(27,34,49,52,66,75,96)$
\\
\hline
$E_8$ & (1,3,5,5,7,7,9,11) & (2,2,3,3,4,4,5,6) & (29,46,57,68,84,91,110,135)&
(29,46,57,68,84,91,110,135)\\
\hline
$H_3$ & (1,2,3) & --- && \\
\hline
$H_4$ & (1,5,8,12) & --- && \\
\hline
$I_2(k)$ & $(1,k)$ & --- && \\
\hline
\end{tabular}
\end{table}

\item
Any algebraic Hamiltonian $h$ of a~trigonometric model preserves one or several f\/lags of invariant subspaces with
characteristic vectors given by the highest root vector, the Weyl vector and the Weyl co-vector (see
Table~\ref{table1}). With the single exception of the $E_8$ case the f\/lags for rational and trigonometric models
coincide.

\item
The original Weyl-invariant periodic Hamiltonian~\eqref{H} written in the fundamental trigonometric
invariants~\eqref{INV} corresponds to a~particle moving in f\/lat space with (trigonometric) Arnold metric ${\cal A}$ in
a rational potential,
\begin{gather*}
{\cal H}(\ta)=-\De_{\cal A}+\sum^{\ell}_k g_k V_k(\ta),
\end{gather*}
where $\De_{\cal A}$ is the Laplace--Beltrami operator, $g_k$, $k=1,\ldots,\ell$ are coupling constants, $\ell$ is the
number of dif\/ferent root lengths in the root space. $V_k(\ta)$
are rational functions. So far, we are unaware about the explicit form of the functions $V_k(\ta)$ for all root systems
except for some particular cases (see~\eqref{HBC1-tau}, \eqref{HBC1-tau-qes}, \eqref{HBC2-tau}, \eqref{HBC3-tau}).

\item
The existence of an algebraic form of the Hamiltonian $h$ of a~trigonometric model allows us to construct integrable
discrete systems in the space of orbits with the same hidden algebra structure, having a~property of isospectrality, on
uniform, exponential and mixed uniform-exponential lattices following the strategy presented in~\cite{st:1995} (uniform
lattice) and~\cite{ct:2001} (exponential lattice).

\item The space of orbits formalism allowed us to show that both rational and trigonometric models for any root system
are essentially algebraic: the (appropriately) gauge-rotated Hamiltonians are algebraic operators, their invariant
subspaces are spaces of polynomials. A natural question to ask is: How the elliptic Calogero--Moser systems look like
in a~space of orbits formalism; are they algebraic just like rational and trigonometric systems? A main obstruction to
get an answer is that, in general, it is not known how to construct elliptic invariants~-- the invariants with respect
to a~``double''-af\/f\/ine Weyl group (the Weyl group plus two translations)~-- on a~regular basis. However, such
invariants can be constructed explicitly for two particular root systems: $A_1/BC_1$~\cite{Turbiner:2000}
and $BC_2$~\cite{Turbiner:2012tt}. It can be shown that the corresponding elliptic systems are algebraic.
\end{itemize}

\subsection*{Acknowledgements}

This work was supported in part by the University Program FENOMEC, by the PAPIIT
grant IN109512 and CONACyT grant 166189~(Mexico).

\pdfbookmark[1]{References}{ref}
\LastPageEnding

\end{document}